\documentclass[10pt]{iopart}

\pdfoutput=1

\usepackage{iopams}

\usepackage{stackengine, tabularx, soul}
\usepackage[caption=false]{subfig}
\usepackage{blindtext,scrextend}
\usepackage{color,xcolor,enumitem,times,bbm,bm,array,multirow}
\usepackage[latin1]{inputenc}
\usepackage[T1]{fontenc}
\usepackage{lmodern}
\usepackage{adjustbox}
\usepackage{cite}
\usepackage{setspace}
\usepackage{dsfont}

\newcommand{\bra}[1]{\left< #1 \right|} 
\newcommand{\ket}[1]{\left| #1 \right>}

\begin{document}

\article[Review]{Review}{Photonic quantum information processing: a review}

\author{Fulvio Flamini, Nicol\`o Spagnolo and Fabio Sciarrino}

\address{Dipartimento di Fisica, Sapienza Universit\`{a} di Roma, Piazzale Aldo Moro 5, I-00185 Roma, Italy}
\ead{fabio.sciarrino@uniroma1.it}
\vspace{10pt}

\begin{abstract}

Photonic quantum technologies represent a promising platform for several applications, ranging from long-distance communications to the simulation of complex phenomena. Indeed, the advantages offered by single photons do make them the candidate of choice for carrying quantum information in a broad variety of areas with a versatile approach. Furthermore, recent technological advances are now enabling first concrete applications of photonic quantum information processing. The goal of this manuscript is to provide the reader with a comprehensive review of the state of the art in this active field, with a due balance between theoretical, experimental and technological results. When more convenient, we will present significant achievements in tables or in schematic figures, in order to convey a global perspective of the several horizons that fall under the name of photonic quantum information.

\end{abstract}

%
%
%
%
\ioptwocol

\section*{Introduction}

Nearly thirty years ago, a short period in the history of modern physics, quantum algorithms were presented to the scientific community as ingenious strategies to realize tasks thought to be hard with classical approaches. Significant advances in this sense sprouted within only a few years from different areas of research in cryptography \cite{Bennett84}, algorithms \cite{Shor94}, simulation \cite{Lloyd96} and communication \cite{Bennett93}, ever since placing a \textit{quantum-} before the name of each field. This shift of paradigm required to think out of the box and to start considering quantum mechanics not only as a stage to investigate but, as much interestingly, as a resource for practical purposes.

Photons naturally fit this quantum revolution \cite{Dowling03} as an effective system to process information: they propagate fast, do not interact with the environment and can be easily manipulated. Moreover, future advances will benefit from the technological infrastructures and know-how already developed in the classical context, thus further encouraging to proceed in this direction. While this aspect is particularly relevant for instance for quantum communication, where the  availability of fiber networks can play a key role, milestone achievements have been reached also in the scope of information processing, raising a bridge between the linear-optical platform and information processing \cite{Knill01}.

Encouraged by these considerations, we will review the state of the art in photonic quantum information to provide the reader with a broad perspective in a unified framework. This review article is structured in three chapters, where we attempt to present the numerous works in a convenient classification, even though most of them easily overlap. Chapter 1 focuses on the single-photon encoding schemes and on the technological state of the art for experimental implementations, namely single-photon sources, integrated circuits and single-photon detectors. Chapter 2 delves into the field of quantum communication, describing various theoretical schemes developed to this purpose, overviewing the developments of quantum repeaters and distributed blind quantum computing protocols for photonic quantum networks. This chapter also presents recent significant achievements in long-distance communication and quantum key distribution. Finally, Chapter 3 presents the latest achievements in photonic quantum simulation, discussing the potentialities of single- and multi-photon quantum walks for quantum computing and simulation in the domain of quantum chemistry.
For the sake of clarity and completeness, we will also refer the interested reader to more specialized literature and review articles on each topic.


\section{Implementing quantum information\break with single photons}

Quantum information can be encoded in a variety of physical systems, ranging from photonic states, solid state devices, atomic or nuclear spin systems to electrons, Josephson junctions, superconducting devices (see Table \ref{table:qubits}). Photonic states present several advantages with respect to other platforms, due to the lack of interaction with the external environment that thus corresponds to long decoherence times. As we discuss in the next sections, this feature is particularly relevant in applications such as long-distance quantum communications. In this section we will review the encoding of quantum information in single photons with a focus on discrete-variable approach in the visible domain. For an overview of quantum information processing in the microwave regime, the interested reader can refer to Refs. \cite{Gu17, You11, Xiang13} for specialized, comprehensive reviews on the field, while for continuous-variable quantum information we refer the reader to Ref. \cite{Braunstein14}. Section \ref{sec:encoding} reviews the main degrees of freedom employed to encode qubits or qudits in this physical platform, while Section \ref{sec:phot.tech} summarizes the main recently-developed technological platforms to generate, manipulate and detect single-photon states.

\subsection{Single-photon encoding}
\label{sec:encoding}
\begin{figure}[t!]
\centering
\includegraphics[trim={0 0 0 0},clip, width=\linewidth]{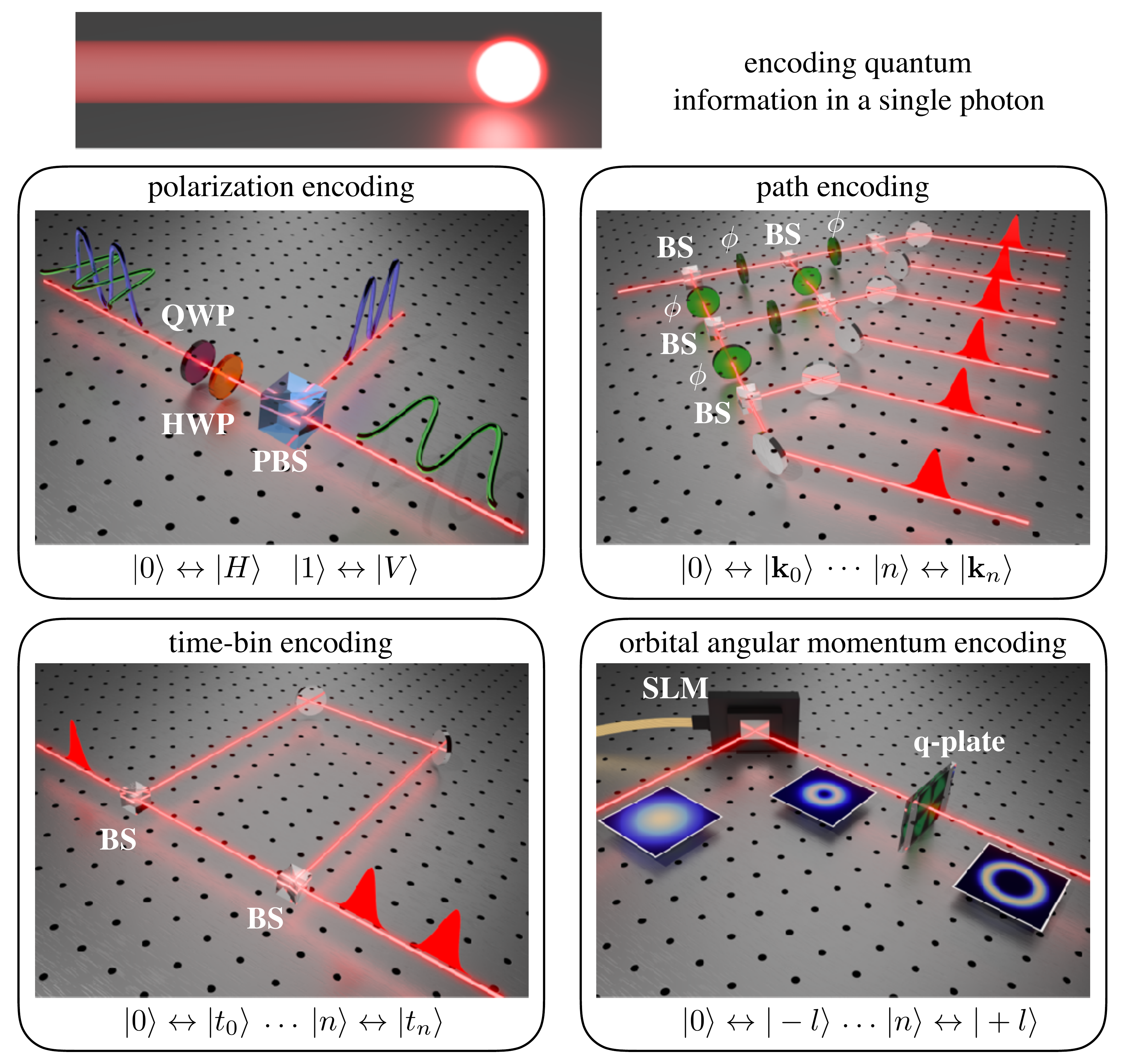}
\caption{\footnotesize Encoding quantum information in a single photon exploiting different degrees of freedom. Possible choices include polarization (only a qubit of information can be carried in this case), path, time-bin and orbital angular momentum (larger dimensionalities can be riched). Legend - QWP: quarter-wave plate, HWP: half-wave plate, PBS: polarizing beam splitter, BS: beam splitter, $\phi$: phase shift, SLM: spatial light modulator.}
\label{fig:FigureEncoding}
\end{figure}

Photon-based quantum information uses the degrees of freedom of light (see Fig. \ref{fig:FigureEncoding}), suitable quantities related to propagation directions (\textit{path} encoding), to momentum (\textit{polarization} encoding), to light spatial distribution (\textit{orbital angular momentum} encoding) and to time (\textit{time-bin} and \textit{time-frequency} encoding). All encoding strategies present their own advantages and weaknesses, and can be combined in a hybrid configuration. In this section we will review each of them focusing on discrete-variable systems, briefly recalling their operation with examples from the latest literature.

\begin{table*}[h!]
	\renewcommand*{\arraystretch}{1.2}
	\footnotesize                  
	\centering
	\caption{\label{table:qubits} Physical systems for encoding a qubit. Some degrees of freedom also allow for the implementation of qudits.}
	The choice for the most suitable system for implementing a qubit depends on the task: quantum computation, communication or simulation. In this review we will focus on discrete-variable photonic implementations.
	\begin{center}
		\begin{tabular*}{\textwidth}{c@{\extracolsep{\fill}}cc}
			\mr
				System & Quantity & Encoding \\
			\mr
			\multirow{4}{5cm}{ \centering{Single photon}}
				& Polarization &  Horizontal / Vertical \\
				& Orbital angular momentum &  Left / Right \\
				& Number &  0/1 photons \\
				& Time & Early/Late \\ 		\hline
			\centering{Continuous-variable fields} & Quadratures & Amplitude-/Phase- squeezing \\ \hline
			\multirow{3}{5cm}{\centering{Josephson junction}}
				& Charge & 0/1 Cooper pair \\
				& Current &  Clock-/Counter- clockwise  \\
				& Energy & Ground/Excited state \\ 		\hline
			\centering{Quantum dot, Optical lattice, Nuclear spin} &    Spin & $\uparrow/\downarrow$ \\ \hline
			\multirow{2}{5cm}{\centering Electrons}
				& Charge & 0/1 electrons \\
				& Spin &  $\uparrow/\downarrow$\\ \hline
			\centering{Non-abelian anyons} & Topology & Braiding \\ 
			\mr
		\end{tabular*}
	\end{center}
\end{table*}

\subsubsection{Encoding in angular momentum --}

The angular momentum of light concerns the rotation of the electromagnetic field vector, a dynamic quantity that influences light-matter interaction.
Two forms of angular momentum exist: spin angular momentum (SAM), associated to its circular polarization, and orbital angular momentum (OAM), associated to the spatial structure of the wavefront. While the two mechanisms cannot be separated in the general case of focused or divergent light beams, their nature becomes manifest for sufficiently collimated ones \cite{MartinezHerrero10}. In the latter case, the total angular momentum $\mathbf{J}$ takes the simpler form

\begin{equation}
\label{eq:TotalAngularMomentum}
\mathbf{J}  \propto \int d^3\mathbf{r}\left[ \Big(|E_L|^2 - |E_R|^2 \Big) - \sum_{j:x,y,z} \left(\imath E_j^\ast \frac{\partial E_j}{\partial \phi} \right) \right]
\end{equation}

\noindent where the first term, the SAM contribution, corresponds to the familiar left (\textit{L}) and right (\textit{R}) circular polarizations while the second, associated to the OAM, describes wavefronts with the typical helical profiles and is thus related to light spatial distribution. The paraxial approximation is the natural framework for light manipulation and encoding \cite{Kok10}, thus in the following sections we will focus on this regime. For specialized reviews, we refer the reader to Refs. \cite{Bliokh15_1, Bliokh15_2}.

\paragraph{Polarization.} Polarization qubits represent a common and practical way to encode quantum information for most applications, thanks to the ease of generation (see Section \ref{sec:sources}) and to the availability of effective tools for manipulation. Indeed, in the limit of paraxial waves of Eq.\eref{eq:TotalAngularMomentum}, the SAM component interacts with anisotropic transparent systems such as birefringent crystals, which are consequently widely employed for polarization manipulation.

A qubit encoded in polarization is usually written as $\ket{\Psi} = \alpha \ket{H}+ \beta \ket{V}$, where \textit{H} and \textit{V} stand for horizontal and vertical polarization, respectively, and

\begin{equation}
\ket{p}  = \int_{-\infty}^{\infty} d\mathbf{k} \ f(\mathbf{k})\ e^{-\imath w_k t} \ \hat{a}^\dagger (\mathbf{k},p) \ket{0}
\end{equation}

\noindent where $p=(H,V)$, $f(.)$ is a wave packet mode function and $\hat{a}^\dagger (\mathbf{k},p)$ is the creation operator for a photon with momentum $\mathbf{k}$ and polarization $p$ \cite{Kok10}. Polarization qubits can be as well expressed in the diagonal basis $\ket{\pm} = \frac{1}{\sqrt{2}}\left( \ket{H}\pm \ket{V} \right)$ or with left and right circular polarizations as $\ket{^L/_R} = \frac{1}{\sqrt{2}}\left( \ket{H} \pm \imath \ket{V} \right)$. Together, the three pairs of states form a set of mutually unbiased bases encoded in polarization \cite{Hou15}, at the core of several applications that we will overview in the following sections.

Polarization encoding has always played an important role in a significant number of investigations in quantum information, ranging from quantum simulation \cite{Sansoni12, Matthews13} to quantum computation \cite{Walther05, Prevedel07, Heilmann14, Barz14, Ciampini16} and communication \cite{Kim01, Ma16}. Its ubiquitous presence in quantum information processing has been further enhanced by the numerous advances in entanglement generation, manipulation and distribution \cite{Wang16, Valles13, Sansoni10, Olislager13, Muller14, Matsuda12, Kaiser14, Hamel14, Bhatti15, Barreiro10} and to the development of suitable theoretical and experimental frameworks for its manipulation in integrated devices \cite{Crespi12, RojasRojas14, Bonneau12}. Moreover, polarization qubits are increasingly coupled to other degrees of freedom of single photons, such as orbital angular momentum \cite{Bhatti15, Barreiro10, Fickler12, Nagali09prl}, path \cite{Walther05, Prevedel07, Ciampini16, Vallone09, Orieux15, Wang16}, time-energy \cite{Chen06, Steinlechner17} and all together \cite{Barreiro05} in so-called hyper-entangled states \cite{Kwiat97}, efficient resources for quantum computation, communication and work extraction protocols.

\paragraph{Orbital angular momentum.}

The OAM component, related to the spatial distribution of the electromagnetic field, can be in turn divided in two terms: one \textit{internal}, origin-independent and associated to twisted wavefronts, and one \textit{external}, given by $\mathbf{L}_e=\mathbf{r} \times \mathbf{P}$ and thus origin-dependent. Quantum information processing based on OAM refers to the first component. The OAM carried by such \textit{optical vortices} is described by the phase term $e^{-\imath q \theta}$, where $\theta$ is the angular coordinate and $q$ is an unbounded integer. For eigenstates of the orbital angular momentum we have $q=Q=\frac{1}{2\pi} \oint d \xi$, where the integral is evaluated around the vortex singularity for a field with phase $\xi$, and $Q$ is the \textit{topological charge} that counts the number of helices in the phase profile. Correspondingly, single photons will carry quantized values of OAM given by $L=q \hbar$ \cite{Cardano12}.

Different techniques have been developed so far to produce and manipulate OAM states. An efficient tool is provided by spiral wave plates \cite{Schemmel14, Bierdz13}, whose thickness increases in a transparent spiral structure so that light experiences a phase gradient during the propagation. However, their cost, the selectivity in wavelength and the hardness to generate qudits still limit their applicability \cite{DAmbrosio16}. Further widely employed tools are metamaterials or cylindrical lens pairs \cite{Willner15}, capable to convert Hermite-Gauss to Laguerre-Gauss OAM modes, diffraction gratings in the form of pitch-fork holograms \cite{Cardano15, Gruneisen11} or spatial light modulators (SLM) \cite{Lavery12, Cardano15, Cardano16}, capable to modify the intensity and/or the phase profile of a beam point-by-point. The ease and versatility of SLMs can be exploited in many applications ranging from optical communication \cite{Krenn14, Bolduc13} to holography-based optical tweezers \cite{Padgett11, Dholakia11}.
Lastly, interesting possibilities are offered by the \textit{q-plate} (QP), a device built on liquid crystals, polymers or sub-wavelength gratings that allows to manipulate OAM depending on the input polarization state \cite{Karimi09, Cardano12, Cardano15, Cardano16, Zhang10, Giovannini11, Nape17, D'Errico17, D'Ambrosio12, D'Ambrosio13, Nagali09}. In the so-called \textit{tuning} condition, a $q$-plate implements the transformations $QP \ket{^L/_R}\ket{l} =  \ket{^L/_R}\ket{l \pm 2q}$, i.e. it flips the qubit polarization in the circular basis and shifts the OAM of a quantity $\Delta l=\pm 2q$. Here, $q$ is related to the topological charge of the QP, which in turn depends on its internal pattern. Applications offered by the QP \cite{D'Ambrosio12, D'Ambrosio13, D'Ambrosio12alignment, DAmbrosio16, Nagali09prl} include the generation of intra-photon entanglement between polarization and OAM degrees of freedom by producing states of the form 
\begin{equation}
QP \ket{_{H}^{V}}\ket{l} =  \ket{_{L}^{R}}\ket{l \pm 2q} \pm  \ket{_{R}^{L}}\ket{l \mp 2q}.
\end{equation}

An important aspect for OAM encoding is to develop practical and reliable techniques to analyze it. While, in fact, polarization encoding requires to resolve only two components, which can be done easily by means of waveplates and polarizing beam splitters, the number of OAM modes is potentially unbounded and as such challenging to characterize \cite{Giovannini13}. Several methods exist so far (see Ref. \cite{D'Errico17} and references therein): the above-mentioned spiral phase plates \cite{Schemmel14, Bierdz13}, holograms and spatial light modulators \cite{Lavery12, Cardano15, Cardano16, Gruneisen11} and q-plates \cite{Karimi09, Cardano12, Cardano15, Cardano16, Zhang10, Giovannini11, Nape17, D'Errico17, D'Ambrosio12, Nagali09},  as well as diffractions through apertures, interference with uniform plane waves, Dove prism interferometers, rotational Doppler frequency shifts and spatial sorting of helical modes \cite{Bhatti15, Karimi09}. In this direction, a spectrum analyzer is a device capable to measure the instantaneous power and phase distributions of OAM components \cite{Zhou17oam, Gruneisen11, Bierdz13, Mirhosseini13, Malik14, Forbes16, Zhao17oam, Piccirillo15}. Once generated and analyzed, OAM states can be manipulated with high degree of control to encode quantum information in the infinite OAM Hilbert space spanned by $l \in Z$, i.e. in the qudits $\psi = \sum_l \alpha_l \ket{l}$ \cite{Barreiro10, Pan16oam, Zhou16prl, Beltran17, Nape17, Babazadeh17, D'Ambrosio12, Nagali09, Zhang14oam, Wang17sagnac}.

OAM states represent a fundamental resource for several applications in quantum information. At a fundamental level, OAM has enabled researches on optimal quantum cloning of OAM-encoded qubits \cite{Nagali09, Bouchard17} and photonic quantum walks in the orbital angular momentum states, with theoretical analyses \cite{Hamilton11, Innocenti17} and experimental demonstrations \cite{Cardano15, Cardano16, Zhang10}. Additional features include the capability of performing a coined quantum walk in the OAM space, providing a viable alternative to other bulk or integrated implementations (see Section \ref{sec:QW}).
Long-distance quantum communication (see Section \ref{sec:LDQC}) benefits from the availability of photonic information carriers able to distribute multiple superposition states in OAM entangled \cite{Ibrahim13, Goyal14teleport, Goyal14qkd, Bhatti15, Krenn15, Hiesmayr16, Malik16, Erhard17, Fickler16, Babazadeh17, Erhard17ghz, Leonhard17, D'Ambrosio12alignment} or hyperentangled \cite{Barreiro10, Giovannini11, Jabir17qkd, D'Ambrosio12, D'Ambrosio12alignment, Farias15, DAmbrosio16, Nagali09prl} states. Moreover, photonic 'flying' qubits encoded in OAM can be prepared in alignment-free states \cite{D'Ambrosio12alignment, Farias15}, being insensitive to rotation of the reference frame. These features have unlocked a number of advancements in free-space quantum key distribution and quantum communication \cite{Vallone14, Mirhosseini15, Bouchard17, Lei15, Wei17, Wang17oam, Pan17fibonacci, Sit17, Goyal14qkd, Mafu13, D'Ambrosio12alignment} with high bit exchange rates \cite{Bozinovic13, Wang12oam}. Related to the capability of producing alignment-free states, hybrid encoding between polarization and OAM can be exploited to enhance the sensitivity to angular rotations \cite{D'Ambrosio13, Jha11}. The confidence in its potential has led to several investigations to support OAM-based photonic quantum networks for delivering information, from sorting \cite{Wei17, Zhang14oam, Wang17oam, Karimi09, Fickler14} and routing OAM states \cite{Garcia-Escartin12, Erhard17, Lavery11} to implementing quantum repeaters with teleportation \cite{Wang15teleport, Goyal13, Goyal14teleport} and quantum memories \cite{Ding13oam, Bussieres14}. First results towards integration of OAM devices have also been reported \cite{Cai12}.
The feasibility of this approach for free-space communication has been further supported by a number of theoretical \cite{Roux11, GonzalezAlonso13, GonzalezAlonso16, Leonhard17, Padgett15} and experimental \cite{Ibrahim13, Farias15, Krenn15} investigations in non-optimal conditions, addressing the issue of beam propagation in a turbulent atmosphere.

\subsubsection{Encoding in propagation direction --}

Path encoding, or encoding in the optical modes in the case of single photons, is the representation of qubits in terms of occupied spatial modes. Path-encoded qubits, as well as qudits \cite{Rossi09}, are perfectly fit for photonic integrated circuits, since waveguide arrays inherently implement a spatial separation and coupling between modes is easily accomplished with directional couplers. Moreover, the high stability and interferometric complexity offered by integrated circuits are useful features for most applications in quantum technologies \cite{Matthews09, Shadbolt12, Ciampini16}. Indeed, accurate control over these qubits for universal manipulation is possible with relatively low technological requirements when compared to other encoding schemes \cite{Bonneau12, Carolan15}. Conversely, an integrated manipulation of polarization qubits requires ad-hoc designs that represent additional challenges, as demonstrated for instance by the fabrication of rotated waveplates in integrated femtosecond laser-written circuits \cite{Corrielli14, Heilmann14}. To this aim, several experimental demonstrations have been shown for tunable all-optical path-entanglement generation via parametric down-conversion in integrated non-linear waveguides  \cite{Silverstone14, Jin14, Harris14, Titchener15}, mostly on SoI and $\textup{LiNbO}_3$ \cite{Solntsev14}. Fiber integrated sources have also been reported for the generation of high-dimensional path-encoded qubits, as interface or preliminary step towards the integration in a circuit \cite{Schaeff12}. Finally, control over non-ideal implementation has been strengthened with investigations on the effect of losses over path-entanglement \cite{Antonosyan14} and on state tomographies on tunable integrated circuits \cite{Shadbolt12}.

\subsubsection{Encoding in time --}

Using time as degree of freedom offers various advantages over other encoding schemes. We present an overview of the approach and of its latest achievements, with a distinction between time-bin and time-energy encodings.

\paragraph{Time-bin encoding.} Time is a natural and effective resource to write information on single-photon quantum states. The mechanism to encode information involves a Mach-Zehnder interferometer with one arm longer than the other. The amplitude associated to an incoming photon is split at the first beam splitter of the Mach-Zehnder and passes through the unbalanced arms: we denote with $\ket{l}$ the state of a photon that has taken the long path, while $\ket{s}$ represents a photon that has taken the short one. This path difference must be stable, i.e. any fluctuation in the temporal delay must be smaller than one wavelength, and longer than the coherence length of each photon to allow a reliable discrimination of the arrival times. A proper dynamic control of the temporal delay is thus desirable to compensate for mechanic and thermal instabilities. A qubit encoded in the photon arrival time can then be written in the superposition $\ket{\Psi} = \frac{1}{\sqrt{2}} \left( \ket{l}+ \ket{s} \right)$, where the states $\ket{p}$ are given by

\begin{equation}
\ket{p}  = \int_{-\infty}^{\infty} dz\ f\left(\frac{t -\frac{z}{c} + p \tau}{\delta t}\right) e^{-\imath \omega (t-\frac{L}{c} + p \tau)} \ \hat{a}^\dagger \ket{0}
\end{equation}

\noindent being $p=(l,s)$, $f(.)$ a wave packet mode function, $\omega$ a fixed angular frequency and $\tau$ the time delay experienced between the two arms of length $L$.

Time-bin encoding presents some advantages with respect to the previous schemes. First, it is suitable for integrated photonic devices, where photons can be generated, manipulated and measured without the need for external encoding devices. Moreover, its resilience to noise acting on polarization, such as depolarizing media or decoherence and mode dispersion, makes time-bin a good candidate for applications in state teleportation \cite{Marcikic03, deRiedmatten04, Landry07}, quantum communication and quantum key distribution, both free-space and in-fiber \cite{Yu15, Marcikic04, Tang16qkd, Gundogan15}. In this direction, several experiments have been carried out for tests of non-locality \cite{Marcikic04, Donohue13, Marcikic02, Guo17timebin}. Entangled photon pairs have been reported with femtosecond pulses \cite{Marcikic02}, including deterministic and narrowband atom-cavity sources \cite{NisbetJones13}, sources integrated on type-II periodically poled lithium niobate \cite{Martin13} and silicon wire \cite{Harada08} waveguides or micro-ring resonators \cite{Wakabayashi15}.  Time-bin manipulation on chip \cite{Xiong15}, storage \cite{Gundogan15} and measurement \cite{Donohue13} have been reported to provide the necessary components for linear-optical quantum networks. Besides quantum communication, time-bin encoding has been proposed as a suitable scheme also for quantum walks  \cite{Schreiber10, Schreiber12, Boutari16} and for photonic \textsc{BosonSampling} \cite{He17, Motes14timebin} (see Section \ref{sec:BosonSampling}).

\paragraph{Time-frequency/energy encoding.} Starting from time-bin encoding, where the allowed states are 'early' and 'late', an extension of this scheme to multiple states consists in correlating the arrival time with single-photons' energy \cite{Zhong15, Nunn13, Steinlechner17}. Relevant applications can be found in quantum communication and quantum key distribution to take advantage of the low decoherence with qubits delivery \cite{Hayat12, Roslund14, Brendel99, Zhong15, Kaiser16, Nunn13, Steinlechner17}. Notwithstanding, frequency-encoded  experimental demonstrations have been reported also in the context of quantum computation \cite{Humphreys13, Campbell14, Menicucci10, Menicucci11, Yokoyama13, Chen14, Soudagar07} often operating with cluster states. Extensions to more than two photons have also been reported \cite{Shalm13}, representing a proof-of-principle investigation for multi-photon Franson interferometry and for engineering discrete- and continuous-variable hyperentangled states.
We conclude this paragraph mentioning theoretical and experimental investigations to manipulate photonic qubits encoded in time \cite{Hosseini09, Soudagar07, Donohue13} and time-frequency \cite{Autebert16, Reddy14, Brecht14, Saglamyurek14, Hayat12, Huntington04, Olislager10}.

\begin{table}[ht!]
	\renewcommand*{\arraystretch}{1.1}        
	\centering
	\caption{\label{SourcesTable} Platforms reported in the last three years for single-photon sources in the range 750-1550 nm.}
	\footnotesize
	\begin{tabular*}{\linewidth}{l@{\extracolsep{\fill}}clc}
		\br
		Process				& Ref.    					& Platform						& $\lambda (nm)$ 		 \\
		\mr
		PDC			 		& \cite{Tian16} 				& PPKTP 						& 795+795  			\\
				 			& \cite{Jabir17source} 		& PPKTP 						& 810+810				\\
		 					& \cite{Weston16} 			& PPKTP 						& 1570+1570	\\
		 					& \cite{Kaneda16} 			& PPKTP 						& \;\,800+1590  			\\
		 					& \cite{Slussarenko17} 		& PPKTP 						& 1550+1550	\\
		 					& \cite{Vergyris16} 			& PPLN 						& 1310+1560  			\\
							& \cite{Krapick16} 			& PPLN 						& 1551+1611+1625		\\
							& \cite{Montaut17}  			& PPLN 						& 1560+1560  			\\
		 					& \cite{Vergyris17} 			& PPLN 						& 1560+1560 			\\
		 					& \cite{Ding15} 				& PPLN						& 1551+1571  			\\
		 					& \cite{Autebert16} 			& AlGaAs  						& 1566+1566			\\
		 					& \cite{Setzpfandt15}	 		& $\textup{LiNbO}_3$				& 1342+1342 			\\
							& \cite{Guo17}				& AlN $\mu R$			 	& 1550+1550			\\  \hline					FWM			 		& \cite{Kultavewuti17}		& AlGaAs 						& 1533+1577 			\\
		 					& \cite{Ramelow15}   		& $\textup{Si}_3\textup{N}_4$ $\mu R$ 		& 1550+1550  	\\
		 					& \cite{Cruz-Delgado16} 		& Few-mode fiber 				& 620+777  			\\
				 			& \cite{Rogers16}			& Silicon microdisk 				& 1497+1534  			\\
				 			& \cite{Cordier17}	 		& LF-HC-PCF		 			& 1552+1552			\\
							& \cite{Yana15}				& Silica FLWw		 			& \;\,830+1130  			\\ \hline
		QD exciton				& \cite{Olbrich17}	 		& InGaAs 						& 1550 			\\
							& \cite{Portalupi15} 		 	& InGaAs 			 			& 945  				\\
							& \cite{Somaschi16} 		 	& InGaAs   					& 890 				\\
							& \cite{Loredo16} 			& InGaAs 		 				& 932  				\\
							& \cite{Kirsanske17} 			& InGaAs 						& 907 		   		\\ 
							& \cite{Schlehahn18} 			& InAs/GaAs 					& 925 				\\
							& \cite{Snijders17} 			& InAs/GaAs 					& 933 			 	\\
							& \cite{Ding16QD} 			& InAs/GaAs 					& 897  				\\
							& \cite{Davanco17} 			& InAs/GaAs 					& 1130  			   	\\ 
		QD biexciton \;			& \cite{Heindel17} 			& InGaAs 						& 905+905 	  		\\
							& \cite{Huber17} 			& GaAs 						& 786+786 		        \\
							& \cite{Jons17} 				& InAsP 						& 930+930				\\
		QD triexciton			& \cite{Khoshnegar17} 		& InAsP 	 					& 894+940		 		\\ \hline
		Fluorescence			& \cite{Benedikter17} 		& Si-V center					& $\sim$750 				\\

		\br
	\end{tabular*}\\
	PDC: spontaneous parametric down-conversion; FWM: four-wave mixing; FLWw: femtosecond laser written waveguide; QD: quantum dot. PPLN: periodically poled lithium niobate. PPKTP: periodically poled potassium titanyl phosphate.\\ InAs: indium arsenide; GaAs: gallium arsenide; InGaAs: indium gallium arsenide; InAsP: indium arsenic phosphide; AlGaAs: aluminium gallium arsenide. AlN: aluminum nitride; Si-V: silicon-vacancy in diamond; LF-HC-PCF: liquid-filled hollow-core photonic crystal-fiber. $\mu R$: microresonator.

\end{table}
\normalsize

\subsection{Photonic technologies}
\label{sec:phot.tech}

\begin{figure*}[t!h]
\centering
\includegraphics[trim={0 0 0 0},clip, width=\textwidth]{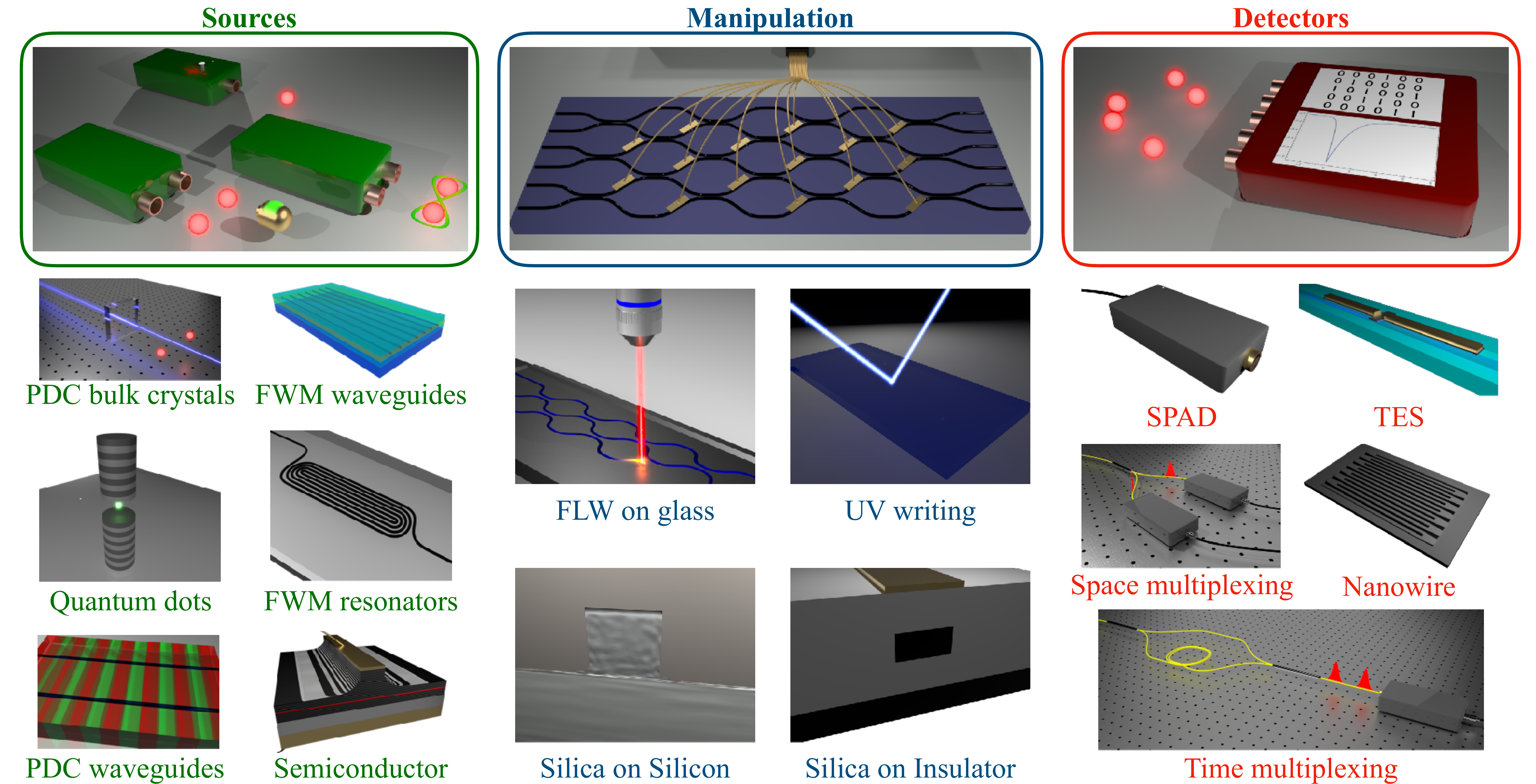}
\caption{\footnotesize Technologies for photonic quantum information processing. Three main stages can be identified. (i) Generation of photonic states, either indistinguishable single photons or entangled states. (ii) Manipulation, where integrated platforms enable apparatuses of increasing complexity. (iii) Measurement of photonic states, where detectors either with photon number resolution or without are currently under development. Legend - PDC: parametric down-conversion, FWM: four-wave mixing, FLW: femtosecond laser writing, SPAD: single-photon avalanche photodiode, TES: transition edge sensor.}
\label{fig:FigureTechnology}
\end{figure*}

In this section we will overview the main technological components for photonic quantum information processing. Three main stages can be identified, as shown schematically in Fig. \ref{fig:FigureTechnology}. First, a crucial requirement is the capability of efficiently generating single-photon states (Sec.\ref{sec:sources}), requiring indistinguishability of correlated states and good control over the degrees of freedom. Then, suitable platforms should be capable of manipulating single- or multi-photon states to perform unitary transformations (Sec.\ref{sec:circuits}). Finally, photons should be efficiently measured with appropriate detection systems (Sec.\ref{sec:detectors}).

\subsubsection{Single-photon sources --}
\label{sec:sources}

Ideally, a good single-photon source should emit only one photon at a time, on demand, at high generation rates and in well-defined states in spatial, temporal and spectral modes. Moreover, different sources should be capable to generate identical photons and their implementation should allow for integration in miniaturized platforms. In parallel, it is crucial to generate correlated states of more than one photon, being entanglement a key resource in several quantum information protocols (see Section \ref{sec:entanglement}). Current sources can fullfil only a limited number of the above requirements, while keeping very high performances for specific applications.

Several approaches for single-photon sources have been developed in the last decades  (see Table \ref{SourcesTable}) \cite{Wang16ten, Ramelow15,Guo17,Setzpfandt15,Weston16,Kaneda16,Vergyris16,Tian16,Montaut17,Jabir17source,Vergyris17,Krapick16,Yana15, Kultavewuti17, Cruz-Delgado16,Rogers16,Higginbottom16,Benedikter17,Peng16,Somaschi16,Loredo16,Heindel17,Olbrich17, Jons17,Khoshnegar17,Ding15,Ding16QD,Davanco17,Schlehahn18,Snijders17,Huber17,Kirsanske17,Portalupi15,Geng16,Li17,Cordier17, Orieux13, Horn13, Spring17, Kruse15, Sansoni17, Atzeni18}. Among probabilistic sources we mention (a) parametric down-conversion (PDC) in bulk crystals \cite{Wang16ten}, semiconductors \cite{Boitier14, Autebert16}, microresonators \cite{Ramelow15, Guo17} and optical waveguides \cite{Setzpfandt15, Weston16,Kaneda16, Vergyris16, Tian16, Montaut17, Jabir17source, Vergyris17, Krapick16, Hamel14, Ding15, Orieux13, Horn13, Kruse15, Sansoni17, Atzeni18} and (b) four-wave mixing (FWM) in optical waveguides \cite{Yana15, Kultavewuti17, Spring17}, few-mode fibers \cite{Cruz-Delgado16} and microdisks \cite{Rogers16}. Deterministic sources include trapped ions \cite{Higginbottom16}, colour centers \cite{Benedikter17} and quantum dots (QD) \cite{Lodahl17} with GaAs \cite{Huber18}, InGaAs \cite{Muller14,Portalupi15, Somaschi16, Loredo16, Heindel17}, InAsP NW \cite{Jons17, Khoshnegar17} or InAs/GaAs \cite{Ding16QD, Davanco17,  Schlehahn18, Snijders17}.

\begin{table*}[h]
	\renewcommand*{\arraystretch}{1.1}  
	\centering
	\caption{\label{CircuitsTable} Integrated circuits for quantum information processing.}
	\footnotesize
	\begin{tabular*}{\textwidth}{c@{\extracolsep{\fill}}ccccccc}
		\br
		Year	 & Ref. 			& Technology 	& Photons & Modes 	& $\lambda (nm)$ & Tunable $\phi$ & Application \\
		\mr
		2008	 & \cite{Politi08} 	& SoS  		& 2 		& 6		& 804 	& n.r. 		& CNOT \\
		2009	 & \cite{Politi09} 	& SoS 		& 4 		& 12 		& 790 	& n.r. 		& Shor's factoring algorithm \\
			 & \cite{Smith09} 	& UVW, SoS	& 2 		& 2 		& 830 	& 1 			& First reconfigurable UVW circuit  \\
			 & \cite{Matthews09} & SoS 		& 4 		& 2 		& 780 	& 1 			& First reconfigurable SoS circuit  \\ 
		2010	 & \cite{Laing10} 	& SoS 		& 2 		& 6 		& 804 	& n.r. 		& CNOT \\
			 & \cite{Sansoni10} 	& FLW 		& 2 		& 2 		& 806 	& n.r. 		& Polarization insensitive BS \\
		2011	 & \cite{Crespi11} 	& FLW 		& 2 		& 4 		& 806 	& n.r. 		& CNOT with partially polarizing BS \\
		2012	 & \cite{Shadbolt12}	& SoS 		& 2		& 6 		& 808 	& 8 			& State generation/detection, Bell test \\
			 & \cite{Bonneau12} 	& Ti:LN 		& 2 		& 2 		& 1550 	& 1 $^{(a)}$ 	& First reconfigurable Ti:LN circuit  \\
			 & \cite{Bonneau12silicon} & SoI  	& 2 		& 2 		& 1550 	& 1 			& First reconfigurable SoI circuit \\
		2014	 & \cite{Heilmann14} 	& FLW  		& 1 		& 2		& 815	& n.r. 		& Integrated waveplates \\
			 & \cite{Corrielli14} 	& FLW  		& 2 		& 6		& 800 	& n.r. 		& Integrated waveplates \\
			 & \cite{Silverstone14} & SoI  		& 2 		& 2 		& 1550	& 1 			& On-chip interference of two sources \\
			 & \cite{Jin14} 		& LN  		& 2 		& 4 		& 1560 	& 1 			& On-chip interference of two sources \\
			 & \cite{Humphreys14} 	& FLW	& 2 		& 2		& 830	& 1 			& Strain-optic active control \\
			 & \cite{Metcalf14} 	& UVW, SoS  	& 3 		& 6 		& 830 	& 1 			& Teleportation \\
			 & \cite{Peruzzo14} 	& SoS  		& 2 		& 6 		& n.a.	& 8			& Variational eigenvalue solver \\
		2015	 & \cite{Flamini15} 	& FLW  		& 2 		& 2 		& 1550 	& 1 			& First reconfigurable FLW circuit \\
			 & \cite{Carolan15} 	& SoS  		& 3 		& 6 		& 808 	& 30 			& H, CNOT, $\sigma, R(\alpha)$\\
			 & \cite{Xiong15} 	& $\textup{Si}_3\textup{N}_4$ & 2 & 5 & 1550 & 15 		& Time-bin entanglement\\
		2016  & \cite{Ma16} 		& Si  		& Attenuated laser 		& 3 		& 1550 	& 4 $^{(a,b)}$ 	& BB84 protocol \\
			 & \cite{Wang16}	& SoI		& 2		& 4+2 $^{(d)}$	& 1550	& 5 		& Entanglement distribution \\
			 & \cite{Poot16} 	& SiN 		& 2 		& 6 		& 1550 	& n.r. 		& CNOT \\
			 & \cite{Sibson17} 	& SoI 		& Attenuated laser	& 3		& 1550 	& 5 $^{(c)}$	& QKD protocol \\
		2017  & \cite{Ding17} 	& SoI 		& Attenuated laser	& 4+8 $^{(d)}$	& 1550 	& 8+10 $^{(d)}$	& QKD protocol \\
			 & \cite{Harris17} 	& SoI  		& Attenuated laser	& 26 		& 1570	& 176 		& Quantum transport \\
		\br
	\end{tabular*}\\
	SoS: silica on silicon; SoI: silica on insulator; FLW: femtosecond laser written; UVW: UV written;\\BS: beam splitter; Ti:LN: titanium indiffusion lithium niobate; SiN: Silicon nitride.\\ Modulation in (a) polarization, (b) intensity, (c) pulse. (d): two integrated circuits. n.r.: non-reconfigurable; n.a.: not available.

\end{table*}
\normalsize

\subsubsection{Integrated quantum circuits --}
\label{sec:circuits}

Traditional optical instruments consisted of bulk optical components, unavoidably large and unpractical, more susceptible to problems of stability, scalability and adaptability to different applications. Just like the evolution undergone by the electronic components, the last decade has seen a strong effort for an optical miniaturization in dielectric materials \cite{Tanzilli12,Orieux16}.
Integrated optical circuits are built upon architectures of directional couplers, with additional geometries to account for deformation-induced phase shifts. Furthermore, recent developments have shown the capability to introduce active reconfigurable elements, thus allowing the fabrication of multi-purpose devices \cite{Carolan15, Flamini15, Harris17}. One widely adopted material for the integration remains fused silica thanks to its numerous benefits: low propagation losses, low birefringence, operation from visible to infrared, good coupling efficiency with single mode fibers and low temperature dependence. Notwithstanding, and differently from miniaturized electronic circuits, linear optical circuits do not have unique platform and manifacturing technique (see Table \ref{CircuitsTable}) \cite{Bogdanov17}:

\paragraph{Silicon-on-Insulator (SoI).} Si-based platforms, including silicon (Si), silicon nitride (SiN) and silicon carbide (SiC), are advanced platforms whose development benefits from the know-how given by electronics technologies \cite{Vivien13}. Si-based devices present a very high refractive index that allows for reduced-size circuits and that is suitable for nonlinear processes. Limitations are the low mode-matching with optical fibers and the relatively high propagation losses. Slightly more favorable conditions can be met with devices based on III-V compound semiconductors, which include indium phosphide \cite{Abellan16}, gallium arsenide \cite{Wang14} and gallium nitride \cite{Xiong11}.
\paragraph{Silica-on-Silicon (SoS).}  A crystal silicon substrate is covered by a layer of silica ($\textup{SiO}_2$), wherein waveguides are etched with rectangular cross sections. A second layer of undoped silica is then laid on top of the structure to enclose the doped silica core and protect the waveguides \cite{Politi08}. Limitations of the SoS technique are the need for a mask and the restriction to one polarization, due to the birefringence induced by the rectangular cross-sections.
\paragraph{UV writing.} Waveguides are inscribed by focusing a strong laser pulse in a photosensitive B- and Ge-doped silica layer, placed within two layers of undoped silica and on top of a third translating silicon layer \cite{Svalgaard94, Spring13, Metcalf14}. This technique does not require the adoption of masks, thus reducing the complexity of fabrication, and allows for arbitrary 3D geometries.
\paragraph{Femtosecond laser writing (FLW).}  The mechanism underlying the process is the non-linear absorption of strong pulses tightly focused in a glass substrate, which causes a permanent and localized modification in the refractive index. Waveguides are drawn by translating the sample at constant speed in 3D geometries \cite{Crespi16, Chaboyer15}. The possibility of writing circular cross-sections and the low birefringence of silica allow for waveguides with low dependance on polarization \cite{RojasRojas14,Sansoni12, Corrielli14}.

\paragraph \noindent On the above platforms, photonic circuits can be implemented \cite{Burgwal17, Flamini17} according to linear-optical interferometric schemes capable to perform arbitrary unitary evolutions \cite{Reck, Clements16}, or designs optimized for Fourier and Hadamard transformations \cite{Crespi16, Flamini17}.

\subsubsection{Single-photon detectors --}
\label{sec:detectors}

Photodetectors are devices that trigger a macroscopic electric signal when stimulated by \textit{one and only one} incoming photon (photon number resolving detectors, PNR) or by \textit{at least one} photon. Detecting photons with high probability and reliability is a key requirement for most tasks, often representing a bottleneck for the overall efficiency of an apparatus. Due to the very low energy of a single photon ($\sim10^{-19} \,\textup J$), a PNR detector requires high gain and low noise to be able to discriminate the correct number.
Non-PNR detectors include single-photon avalanche photodiodes (SPAD) on InGaAs \cite{Zhang15,Comandar15} or Ge-on-Si \cite{Martinez17,Warburton13}, quantum dots \cite{Weng15}, negative feedback avalanche diodes \cite{Yan12, Korzh14, Covi15}, superconducting nanowires \cite{Li15, Zhang15nw, Yamashita16, Atikian14, Tyler16, Arpaia15, Takesue15,LeJeannic16,Zhang17nbn,Zadeh16,Wang17nbn,Vorobyov17,Miki17,Krapick17}, artificial $\Lambda$-type three-level systems \cite{Inomata16} and up-conversion detectors \cite{Ma17, Pelc11, Hu12, Pelc12,Pomarico10,Zheng16}. Si-based SPADs exhibit good performances with visible light but still suffer in the infrared window, due to the incompatibility between good IR absorption and low-noise. PNR detectors include instead transition-edge sensors \cite{Miller11, Calkins13, Gerrits11, Hopker17, Lamas-Linares13}, parallel superconducting nanowire single-photon detectors \cite{Najafi12, Heath14}, quantum dot coupled resonant tunneling diodes \cite{Weng15}, organic field-effect transistors \cite{Yuan13} and multiplexed SPADs \cite{Avenhaus10, Thomas12}. Recently, large effort has been devoted to the optical integration of superconducting detectors on waveguide structures, such as on LiNb$\textup{O}_3$ \cite{Tanner12,Hopker17}, GaAs \cite{Sprengers11, Jahanmirinejad12, Reithmaier13, Sahin15, Zhou14, Kaniber16, Najafi15, Mattioli16, Li16nw}, Si \cite{Pernice12, Akhlaghi15}, S$\textup{i}_3$N$\textup{i}_4$ \cite{Cavalier11, Ferrari15, Kahl15, Schuck13, Schuck16, Beyer15, Shainline17} and diamond \cite{Rath15, Atikian14}.
For a more detailed discussion on the state of the art, we refer the interested reader to specialized reviews on the topic \cite{Eisaman11, Natarajan12}.

\section{Quantum communication}

Quantum communication aims to connect distant quantum processors with an increased level of security. In recent years, advances in all these areas have led to the implementation of first quantum networks in different locations (see Table \ref{table:qNetworks}). In this section we will briefly overview the state of the art in this research.

\begin{table*}[h!]
	\renewcommand*{\arraystretch}{1.1}  
	\centering
	\caption{\label{table:qNetworks} Worldwide in-fiber photonic quantum networks (not exhaustive list).}
	\footnotesize
	\begin{tabular*}{\textwidth}{c@{\extracolsep{\fill}}ccccc}
		\br
		Network 				& Ref. 					& Launch 	& Location 	& Nodes 	& Distance (km) \\
		\mr
		DARPA 				&\cite{Elliott05} 				& 2003 	& USA 		& 10 		& 29			\\
		SECOQC 				&\cite{Peev09} 				& 2003 	& Austria 		& 6 		& 200 		\\
		SwissQuantum 			&\cite{Stucki11} 			& 2009 	& Switzerland 	& 3 		& 35			\\
		Hierarchical Quantum Network 	&\cite{Xu09, chineseNetwork16} 	& 2009 	& China 	& 32 		& 2000 		\\
		Tokyo QKD Network 		&\cite{Fujiwara16, Sasaki11} 	& 2010 	& Japan		& 5		& 45 			\\
		Los Alamos National Lab 	&\cite{Hughes13} 			& 2011 	& USA 		& n.a. 	& 50 			\\
		\br
	\end{tabular*}\\
	DARPA: Defense Advanced Research Projects Agency; SECOQC: SEcure COmmunication based on Quantum Cryptography. 
\end{table*}
\normalsize

\subsection{Protocols for quantum communication}

Ever since the intuition of Bennett and Brassard with the famous BB84 protocol \cite{Bennett84}, it is known that quantum information allows one to devise algorithms to achieve classically unparalleled results in information transfer. Quantum infrastructures for managing information and delivering entanglement are now believed to join current technologies in a number of relevant tasks. Photons are by far the most suitable physical system to implement flying qubits to deliver information: they experience negligible decoherence through free space or optical fibers, they allow a high spatial control, and the technology for linear-optical devices to manipulate qubits is accessible and at an advanced stage \cite{Nielsen_Chuang, Saleh07}.

\subsubsection{Photons as quantum information carriers --}
\label{sec:entanglement}

\begin{figure*}[t!h]
\centering
\includegraphics[trim={0 0 0 0},clip, width=\textwidth]{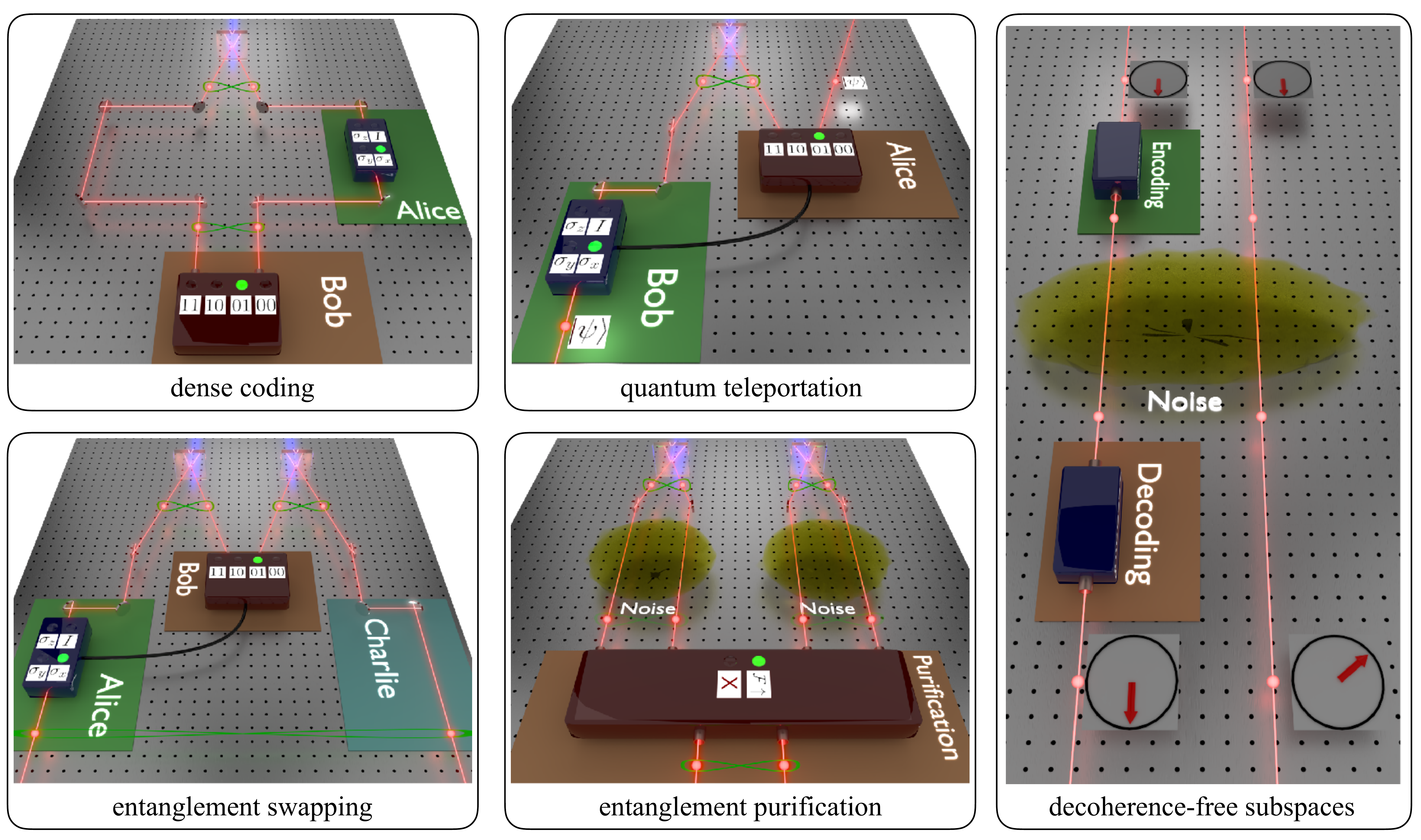}
\caption{\footnotesize Main ingredients for photon-based quantum communication described in this section. Photons represent the most promising system to encode quantum information in this field for their speed, ease of manipulation and long coherence time. }
\label{fig:FigureQuantumCommunication}
\end{figure*}

Quantum communication protocols require two or more channels to exchange information between the parties, at least one classical and one quantum. Proper measurement are carried out on quantum systems shared between the parties, while classical messages can be exchanged to guide or complete the transfer. Entanglement is an essential physical resource to this aim \cite{Horodecki09} and, as such, it is necessary to protect it during the transfer. In the last decades a large number of theoretical investigations was reported to rigorously define and develop the field, which is still growing and reaching now a mature age \cite{Yuan10, Krenn16}. The theoretical framework, however multifaceted and oriented to various improvements, ultimately relies on few fundamental ingredients (see Fig. \ref{fig:FigureQuantumCommunication}) introduced a couple of decades ago, which have now been shown in first demonstrations \cite{Diamanti16, Krenn16}.

\paragraph{Dense coding.} 
Quantum mechanics allows to increase the capacity of a quantum communication channel by transferring two bits of classical information with only one qubit\cite{Bennett92dense}. Suppose Alice and Bob share an entangled photon pair, each keeping one photon. Alice can encode two bits on her photon by choosing and applying on her photon one between four unitary transformations: the identity operation \(\mathds{1}\) (00), a bit flip $\sigma_x$ (01), a phase flip $\sigma_z$ (10) or both a bit flip and a phase flip $\sigma_y$ (11). Alice sends her particle to Bob, who measures both particles in the Bell basis. The outcome of the Bell measurement (BM) will then correspond to the two bits of information sent by Alice.

One challenge for implementing dense conding is represented by the BM stage, since achieving a never-failing BM is impossible using only linear-optical elements such as beam splitters, phase shifters, detectors and ancillary particles \cite{Lutkenhaus99}. A solution to this issue consists in exploiting hyperentangled photons \cite{Barreiro05}, leading to complete BM demonstrations in orbital angular momentum/polarization \cite{Barreiro08}, momentum/polarization \cite{Barbieri07} and time/polarization \cite{Schuck06, Williams17}. The latter schemes have the additional advantage of being feasible to prepare and of enabling an efficient transmission through optical fibers. So far, the maximum channel capacity ($1.665 \pm 0.018$) has been reported in Ref.\cite{Williams17}, beating the limit of $\log_2 3 \sim 1.585$ bits obtainable by means of only linear optics and entanglement, due to the impossibility of perfectly discriminating all four Bell states in this case.

\paragraph{Entanglement purification.}

One requirement is the capability of delivering entanglement without having it spoilt between distant nodes of a network, as it happens with a probability increasing exponentially in the length of a noisy transmission. Entanglement purification allows to circumvent this issue by extracting high-quality entangled pairs from a given ensemble using only local transformations and classical communication \cite{Bennett96}. More specifically \cite{Yuan10}, suppose Alice and Bob share two copies of the mixed state

\begin{equation}
	\rho_{A,B} = \mathcal{F} \ket{\phi^\dagger}_{A,B}\bra{\phi^\dagger}+(1-\mathcal{F}) \ket{\psi^\dagger}_{A,B}\bra{\psi^\dagger},
\end{equation}

\noindent with $\ket{\psi^\dagger} \propto \ket{0,1}+ \ket{1,0} $ and $\ket{\phi^\dagger} \propto \ket{0,0}+ \ket{1,1} $, and they want to purify it so as to increase the probability of having a maximally-entangled state $\ket{\phi^\dagger}$. The original approach was based on CNOT operations \cite{Bennett96}; few years later a simpler solution was introduced \cite{Pan01} and verified experimentally \cite{Pan03} based on only polarizing beam splitters. While the probability of success is $50\%$ lower than the one with CNOT gates, the quality of these optical elements allow for a purification with much higher precision and control. Via a proper combination of projective measurements in polarization, it is possible to increase the amplitude associated to the maximally-entangled pair $\ket{\phi^\dagger}$, which can be used to quantify the fidelity of the transmission, as $\mathcal{F}  \rightarrow  \mathcal{F}^2/\left(\mathcal{F}^2 + (1-\mathcal{F})^2\right)$ when $\mathcal{F}>\frac{1}{2}$.\\ Entanglement purification is a key ingredient also in quantum computation, such as for quantum error-correction protocols, in quantum cryptography and quantum teleportation. For a more complete overview of the state of the art we refer to Refs. \cite{Yuan10,Sheng15,Zhang17,Simon17,Kalb17}.

\paragraph{Quantum state teleportation.}
Ever since its introduction \cite{Bennett93}, quantum state teleportation (QST) has been one of the most notable examples of quantum communication. In terms of classical communication, QST enables the transfer of one qubit by sending two bits of classical information, i.e. a reversed dense coding. The mechanism works as follows: suppose Alice and Bob share an entangled photon pair $\ket{\psi^-}_{A,B} = \frac{1}{\sqrt{2}} \left( \ket{0,1}_{A,B} - \ket{1,0}_{A,B} \right)$ and Alice wants to teleport to Bob the -unknown- qubit $\ket{\psi}_{T} = \frac{1}{\sqrt{2}} \left( \alpha \ket{0}_{T} + \beta \ket{1}_{T} \right)$. The complete system of three photons has the form

\begin{eqnarray}
\ket{\psi}_{T} \otimes \ket{\psi^-}_{A,B} = &-&  \frac{1}{2} \ket{\psi^-}_{T,A} \left( \alpha \ket{0}_{B} + \beta \ket{1}_{B} \right)  \label{qT:a} \\ 
 &-& \frac{1}{2} \ket{\psi^+}_{T,A} \left( \alpha \ket{0}_{B} - \beta \ket{1}_{B} \right)   \label{qT:b} \\
  &+&  \frac{1}{2} \ket{\phi^-}_{T,A} \left( \alpha \ket{1}_{B} + \beta \ket{0}_{B} \right)   \label{qT:c} \\
   &+&  \frac{1}{2} \ket{\phi^+}_{T,A} \left( \alpha \ket{1}_{B} - \beta \ket{0}_{B} \right) \label{qT:d}.
\end{eqnarray}

\noindent Alice can then perform a joint Bell measurement on her two photons $A$ (entangled with Bob's) and $T$, associate two bits to the Bell state found out of the four and send them to Bob via a classical channel. At this point, Bob can simply perform one of four transformations (identity operation for \eref{qT:a}, phase flip $\sigma_z$ for \eref{qT:b}, bit flip $\sigma_x$ for \eref{qT:c} or both bit flip and phase flip $\sigma_y$ for \eref{qT:d}) on his entangled photon $B$ to shape it as the one -unknown- possessed by Alice.

The original work by Bennett \textit{et al.} \cite{Bennett93} has rapidly triggered a large number of investigations \cite{Nielsen_Chuang} for a broad range of applications \cite{Pirandola15}. Among all, teleportation schemes were shown to enable new approaches for universal quantum computation \cite{Gottesman99, Ishizaka08}, in particular as one-way quantum computers \cite{Raussendorf01}.
From the experimental perspective, numerous achievements have been reported on photonic platforms proving the feasibility of the scheme already with state-of-the-art technology. After the first demonstrations in 1997-1998 \cite{Bouwmeester97, Boschi98}, one further proof appeared with the unconditional teleportation of optical coherent states with squeezed-state entanglement \cite{Furusawa98}. Later on, in 2001 Kim \textit{et al.} reported an experimental teleportation where all four states were distinguished in the Bell-state measurement \cite{Kim01}, while Jennewein \textit{et al.} provided a proof of the nonlocality of the process and of entanglement swapping \cite{Jennewein01}. One year later, in 2002 Pan \textit{et al.} were performing four-photon experiments for high-fidelity teleportation \cite{Pan02} and Lombardi \textit{et al.} teleported qubits encoded in vacuum--one-photon states \cite{Lombardi02}. The beginning of the new century saw a true race towards more complex implementations. In 2004, for instance, a single-mode discrete teleportation scheme using a quantum dot single-photon source has been demonstrated \cite{Fattal04}. At the same time, increasing the teleportation distances \cite{Xia17} became an interesting benchmark to assess the feasibility of practical implementations for future quantum networks \cite{Marcikic03, Ursin04, deRiedmatten04, Landry07, Jin10, Ma12, Yin12}. Today the current record for the longest distance is kept by Ren \textit{et al.} who teleported single-photon qubits from a ground observatory to a satellite 1400 km high in atmosphere \cite{Ren17}. Next to the discrete-variable schemes of the first demonstrations, teleportation was also reported on squeezed entangled states \cite{Zhang03, Bowen03, Takei05, Takei05squeezedTele, Yonezawa07, Yukawa08, Lee11}, for which a review is available \cite{Pirandola06}. Indeed, the main reasons behind the increasing interest toward continuous-variable schemes concern practical advantages, since Bell measurements can be realized by means of only passive linear-optical elements and homodyne detection with very high precision. To bridge the gap between discrete and continuous variables, a hybrid approach has been recently proposed and tested \cite{Takeda13}.
Finally, interesting achievements on photonic teleportation have been demonstrated with the first implementation on integrated circuits \cite{Metcalf14}, which may turn useful for future realizations of quantum network nodes, as well as schemes with simultaneous teleportation of multiple degrees of freedom \cite{Wang15teleport} and teleportation of qudits \cite{Goyal14teleport}, as opposed to the conventional two-level discrete approach.

In the above realizations, where photons encode information through all the stages of the protocol, the state of the qubit was transferred from one photon to another. Though the sources of single photons and photon pairs are still spatially close, the process is indeed the core of future long-distance quantum communication as we envisage it today (see Section \ref{sec:LDQC}) \cite{Kimble08, Krenn16}. One further requirement for a reliable and practical quantum communication is the capability of storing information for subsequent uses. The past decade has seen a strong effort directed towards the development of matter-light interfaces as building blocks for quantum computation and communication, where entanglement between single-photon states and atomic ensembles represents an effective solution. In nearly ten years several works have been carried out in this context \cite{Hammerer10}. Single-photon qubits have been teleported on atomic ensembles \cite{Krauter13, Barrett04} such as Caesium atoms \cite{Sherson06} or $^{87}$Rb atoms \cite{Chen08teleport, Nolleke13, Bao12}, on a pair of trapped calcium ions \cite{Riebe04} or diamond \cite{Hou16}. In this sense, single photons can also turn effective as mediators between distant matter qubits, as already reported with single trapped ytterbium ions \cite{Olmschenk09}, quantum dots \cite{Gao13}, rare-earth-ion doped crystals \cite{Bussieres14} or diamond spin qubits \cite{Pfaff14}.

\paragraph{Entanglement swapping.}
In order to increase the distance between nodes of a quantum network without decreasing the quality of the transmission, i.e. the degree of entanglement distributed, it is necessary to develop techniques to protect or restore the state of the encoding photon. Entanglement swapping is a promising solution to this issue \cite{Khalique14} and, nowadays, entanglement distribution includes it as a fundamental routine in quantum repeaters (see Section \ref{qRepeaters}) \cite{Yuan10}. Entanglement swapping is actually very similar to teleportation, the only difference being that the particle to be teleported is part of a second entangled pair. Schematically, given two pairs of entangled particles \textit{A}-\textit{B} and \textit{C}-\textit{D} shared by Alice and Bob, for instance in the Bell state $\ket{\psi^{-}}$, the goal is to convert the whole system to a new entangled pair \textit{A}-\textit{D} by performing a Bell measurement on \textit{B} and \textit{C}:

\begin{eqnarray}
\ket{\psi^{-}}_{A,B} &\ket{\psi^{-}}_{C,D}&  \propto \nonumber \\ &\ket{\psi^{-}}_{A,D}& \ket{\psi^{-}}_{B,C} \;+\; \ket{\psi^{+}}_{A,D}  \ket{\psi^{+}}_{B,C} \;+  \nonumber  \\
&\ket{\phi^{-}}_{A,D}&  \ket{\phi^{-}}_{B,C} \;+\;  \ket{\phi^{+}}_{A,D}  \ket{\phi^{+}}_{B,C}
\end{eqnarray}

\noindent After the Bell measurement on \textit{B}-\textit{C}, and whatever outcome Alice receives, Bob remains with an entangled pair \textit{A}-\textit{D} even though they did not share any entanglement or interaction. Similarly to teleportation, photons offer a convenient solution for an implementation of the scheme.

First demonstrations of entanglement swapping go back to 1998 by Pan \textit{et al.} \cite{Pan98}, to 2001 by Jennewein \textit{et al.} and to 2002 by Sciarrino \textit{et al.}, all based on parametric down-conversion sources. Subsequent experiments still employed probabilistic single-photon sources at near-infrared and telecom wavelengths, the telecommunication windows for optical fibers, in bulk or waveguide as in Ref.\cite{Jin15} and references therein. Entanglement swapping has also been tested with discrete \cite{Sun17, Weston18} and continuous variables \cite{Takei05}, or even with a hybrid approach \cite{Takeda15}.
Future implementations of quantum repeaters in optical networks will require active synchronization of the single-photon sources involved in the exchange of information between the various nodes. To this aim, proof-of-principle demonstrations of this scheme have been reported in the last decade, where two synchronized independent sources generated entangled photon pairs for entanglement swapping and non-locality tests \cite{Yuan10}.

\subsubsection{Decoherence-free communication --}
\label{sec:decoh.free}

So far, we have described how entanglement purification, entanglement swapping, dense coding and teleportation allow for the distribution of entanglement, the essential resource for quantum communiction, between in-principle arbitrarily distant locations. However, non-ideal operating conditions may hinder or completely prevent their concrete implementation. Two methods can be adopted to circumvent the issue of noisy channels: to increase the resilience of a protocol to losses and noise \cite{Gottesman03, Xiang10, Kocsis13} (two-way classical communications), or to decrease their influence by encoding information in immune states (decoherence-free subspaces, DFS), for both quantum computation \cite{Lidar98} and quantum communication tasks \cite{Klein06}. In the following we provide simple examples of DFS schemes, while further on we will overview some of the most recent experimental demonstrations.

In DFS protocols, qubits are encoded in states that do not experience decoherence in a given channel thanks to known symmetries of the subspace.
As an example \cite{Yuan10}, let us consider two qubits evolving through a unitary transformation $U(t)$ such that
$ U(t) \ket{0} = e^{-\imath \omega_0 t} \ket{0}$ and $ U(t) \ket{1} = e^{-\imath \omega_1 t} \ket{0}$. We can see that $\ket{\psi^{+}} = \frac{1}{\sqrt{2}} (\ket{01}+\ket{10}) $ is an invariant of $U(t)$ since $ U(t) \ket{\psi^{+}} = e^{-\imath (\omega_0+\omega_1) t}  \ket{\psi^{+}}$, which is equivalent to $\ket{\psi^{+}}$ up to a global phase.
A second example is given by the unitary transformation

\begin{equation}
	\cases{
	U(\theta, \alpha) \ket{0} = \cos{\theta} \ket{0} + e^{\imath \alpha} \sin{\theta} \ket{1}  \\
	U(\theta, \alpha) \ket{1} = e^{- \imath \alpha} \sin{\theta} \ket{0} + \cos{\theta} \ket{1}
	}
\end{equation}

\noindent The transformation $ U(\theta,0)$ is called \textit{collective rotation noise}. In this case, the states $\ket{\phi^{+}} = \frac{1}{\sqrt{2}} (\ket{00}+\ket{11}) $ and $\ket{\phi^{-}} = \frac{1}{\sqrt{2}} (\ket{01}-\ket{10}) $ are invariants, while $\ket{\phi^{-}}$ remains unchanged even under the more generic $U(\theta,\alpha)$. Thus, by modelling the noise in a given quantum channel, it is in principle possible to construct quantum states that are immune to all decoherence effects corresponding to that specific pattern.

The first experimental demonstration of DFS communication was reported in 2000 by Kwiat \textit{et al.} \cite{Kwiat00}, where the authors induced a controllable collective decoherence on an entangled photon pair generated by spontaneous parametric down-conversion (PDC) and observing that, as predicted, one specific entangled state was immune to decoherence. More recently, in 2005 Jiang \textit{et al.} reported a test of DFS against collective noise on polarization and phase with four two-qubit states again from PDC \cite{Jiang05}, while in 2006 Chen \textit{et al.} used single photons, entangled in polarization and time, to compensate for errors induced by a collective rotation of the polarization \cite{Chen06}. Importantly, their scheme was also alignment-free, i.e. with no need for a shared reference frame, and insensitive to phase fluctuations in the interferometer. In the same year, Zhang \textit{et al.} reported a fault-tolerant implementation of quantum key distribution (see Section \ref{sec:QKD}) with polarization encoding, capable to account for bit-flip errors and collective rotation of the polarization state, without the need for a calibration of the reference frame \cite{zhang06}. In 2008, Yamamoto \textit{et al.} demonstrated an entanglement distribution scheme with state-independent DFS from PDC, with an approach robust against fluctuations of the reference frame between distant nodes \cite{Yamamoto08}. Other significant features reported in their work were the capability to extend the scheme to multipartite states, thanks precisely to the state-independence, and its applicability also with single-mode fibers.

More recently, various proposals and experimental tests have been reported on variants or improvements of the original schemes. For instance, Ikuta \textit{et al.} in 2011 proposed and demonstrated a solution for increasing the efficiency of DFS-based entanglement distribution over lossy channels \cite{Ikuta11}. Additional features of their work were the use of backward propagation of coherent light in combination with single-photon states, and the violation of the Clauser-Horne-Shimony-Holt inequality \cite{Nielsen_Chuang} to prove non-locality in the transmission. In 2017, novel investigations on matter qubits were reported by Wang \textit{et al.}, with a room-temperature quantum memory realized with two nuclear spins coupled to the electronic spin of a single nitrogen-vacancy center in diamond \cite{Wang17dfs}, and by Zwerger \textit{et al.}, with the demonstration of a quantum repeater (see Section \ref{qRepeaters}) using ion-photon entangled states protected via DFS against collective dephasing \cite{Zwerger17}.

\subsection{Long-distance quantum communication}
\label{sec:LDQC}

\begin{figure*}[t!]
\centering
\includegraphics[trim={0 0 0 0},clip, width=\textwidth]{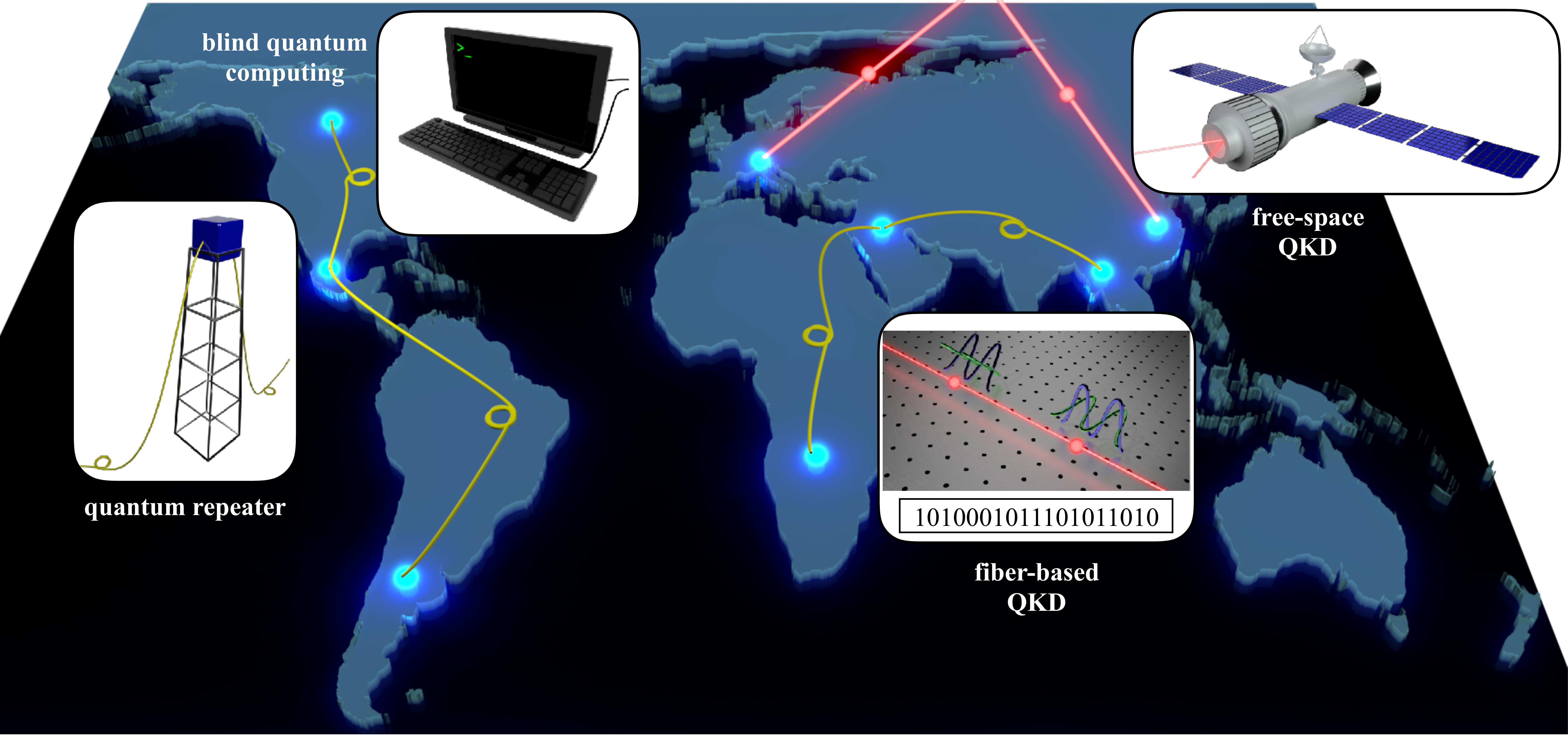}
\caption{\footnotesize Schematic view of the main nodes in a large-scale quantum network, comprising blind quantum computing stages for the end user, quantum repeaters for long-distance transmission and quantum key distribution performed either via fiber networks or via free-space links.}
\label{fig:FigureQuantumNetwork}
\end{figure*}

Future quantum networks, engineered to deliver quantum information between distant nodes on the globe, are the sought-after goal underlying most of the investigations described above (see Fig. \ref{fig:FigureQuantumNetwork}). This challenge already counts first demonstrations in Austria \cite{Peev09}, China \cite{Xu09, chineseNetwork16}, Japan \cite{Fujiwara16, Sasaki11}, Switzerland \cite{Stucki11} and USA \cite{Elliott05, Hughes13} (see Table \ref{table:qNetworks}). We will now overview the state of the art towards their implementations.

\subsubsection{Quantum repeaters --}
\label{qRepeaters}

Quantum communication promises to enable quantum information protocols distributed over distant locations. In the previous section we have shortly analyzed some of the main theoretical ingredients of these quantum networks, together with the first concrete demonstrations of their feasibility. The key resource for its fulfillment is provided by quantum entanglement, at the core of quantum dense coding, teleportation, entanglement purification and entanglement swapping. By using these schemes it is possible to create entanglement between two nodes of a network to transfer quantum information. We have also described how this resource deteriorates easily and exponentially fast with the length of transmission for several reasons. Photon losses are the first cause of deterioration: even though fiber attenuation at telecom wavelength can be partly reduced, even ultra-low loss optical fibers still unavoidably lead to an exponential decrease in the generation rate of entangled pairs. This is a fundamental limitation imposed by quantum mechanics: the maximum number of entanglement bits (ebits) that can be distributed over a lossy channel with transmissivity $T$ is in fact equal to $-\log_2 (1-T)$ ebits per channel use \cite{Pirandola17}. In addition, errors can occurr all along the transmission due to imperfect gates or measurements, without the possibility of duplicating quantum states deterministically \cite{Nielsen_Chuang} and with conditions much more sensitive than in the classical case.
Notwithstanding, two approaches exist to circumvent these issues: reducing the transmission loss using free-space communication \cite{Yin12, Wang12oam, Vallone14, Steinlechner17, SchmittManderbach07, Resch05, Peng05, Jin10, Aspelmeyer03, Wang13satellite, Nauerth13, Ren17, Patel14} and decoherence-free subspaces \cite{Kwiat00,Chen06,Yamamoto08,Zwerger17,Wang17dfs,Ikuta11,Jiang05,GonzalezAlonso13}, or employing quantum repeaters. 

Quantum repeaters (QR), proposed in 1998 by Briegel \textit{et al.} \cite{Briegel98}, currently represent a strong candidate to circumvent these issues. In principle, QRs allow to improve the fidelity of transmission with time-overhead polynomial in the transmission length. In the original proposal, QRs operate by splitting the channel into a suitable number of intermediate segments linked by as many QRs, where active control can compensate for fiber attenuation, gate errors and possible noise. Once a sufficiently strong entanglement is established between two target nodes, routine quantum communication can start effectively. Today numerous schemes and platforms exist for long-distance quantum communication based on QRs \cite{Muralidharan16}. QRs can be grouped in three main classes, according to the technique adopted for correcting propagation errors \cite{Muralidharan16}, as summarized schematically in Table \ref{table:QR}.

\begin{table}[h!] 
	\renewcommand*{\arraystretch}{1.4}                       
	\centering
	\caption{\label{table:QR} Quantum repeaters can be classified in three generations according to the approach used to correct errors.\\Table adapted from Ref.\cite{Muralidharan16}.}
	\footnotesize                 
	\begin{center}
		\begin{tabular*}{\linewidth}{c@{\extracolsep{\fill}}ccccc}
			\mr
			Error			& Solution			& I			&  II			& III \\
			\mr
			\multirow{2}{1cm}{\centering{Loss}}
				& HEG ($\leftrightarrow$) 	& \checkmark 	& \checkmark 	&  \\
				& QEC ($\rightarrow$)	&   			&   			& \checkmark\\ 		\hline
			\multirow{2}{1cm}{\centering{Gate}}
				& HEP ($\leftrightarrow$)	& \checkmark 	&   			&  \\
				& QEC ($\rightarrow$)	&   			& \checkmark 	& \checkmark\\
			\mr
		\end{tabular*}
	\end{center}
	HEG: Heralded entanglement generation. HEP: Heralded entanglement purification. QEC: quantum error correction.\\
	$\longrightarrow$: one-way communication; $\longleftrightarrow$: two-way communication.
\end{table}

The first generation of QRs (I) employs heralded entanglement generation (HEG) and heralded entanglement purification (HEP) to reduce the deterioration induced by losses from an exponential to a polynomial scaling (see Section \ref{sec:entanglement}) \cite{Briegel98, Sangouard11, Dur99, VanLoock06, Duan01, Zwerger12}. The idea here is to create entanglement between adjacent nodes and to use teleportation to exchange information. Entanglement swapping and purification can then be properly combined to sequentially extend the distribution of entangled photon pairs until two nodes are connected. The overall rate for quantum communication based on this scheme decreases polynomially with the distance; however, a good solution to this drawback could be multiplexing in one of the degrees of freedom (see Section \ref{sec:encoding}) to increase the transmission rate. A number of other approaches have been tested experimentally in the last decade. An efficient scheme using double-photon guns based on semiconductor quantum dots, polarizing beam splitters and probabilistic optical CNOT gates was proposed in 2003 by Kok \textit{et al.} \cite{Kok03}. Active purification of arbitrary errors using only two qubits at each QR, based on nuclear and electronic spins in nitrogen-vacancy color centers in diamond, was proposed by Childress \textit{et al.} \cite{Childress06}. Further proposals involve multi-mode memories based on photon echo in solids doped with rare-earth ions \cite{Simon07}. A combination of photonic and atomic platforms was also investigated by Sangouard \textit{et al.} \cite{Sangouard09, Sangouard11}, while combinations of quantum-dot qubits and optical microcavities have been studied by Wang \textit{et al.} \cite{Wang12repeater}. Recently, a novel measurement-based QR was shown to enable entanglement purification and entanglement swapping \cite{Zwerger12} for one- and two-dimensional networks using cluster states and photon-ion entanglement \cite{Wallnofer16, Zwerger17}. A thourough analysis under noisy conditions of measurement-based quantum network coding scheme for QRs was later provided by Matsuo \textit{et al.} \cite{Matsuo17}. Finally, for an overview on quantum memories in the context of quantum communication we refer to Ref.\cite{Simon17} and references therein.

The second generation of QRs (II) employs (i) HEG to reduce the deterioration induced by losses and (ii) quantum error correction (QEC) to correct gate errors  \cite{Jiang09, Munro10}. Teleportation with CNOT gates and entanglement swapping are still sequentially applied to extend the entanglement to distant nodes; however, the use of QEC in place of HEP speeds up the process by avoiding the time delays due to non-adjacent-nodes signalling \cite{Bratzik14}. With this approach, quantum memories are employed at each side of a QR to save the state of the entangled photon pairs while waiting for the classical signals to carry out teleportation \cite{Simon07,Nolleke13,Munro12,Mazurek14,Heshami16,Gundogan15,Ding13oam,Chen08teleport, Campbell14,Bussieres14,Bao12,Laplane17, Kalb17}. Indeed, QEC with qubit-repetition codes and Calderbank-Shor-Steane codes \cite{Nielsen_Chuang} were shown to be effective against imperfect quantum memories operation, photon losses and gate errors \cite{Bernardes12}. However, starting from 2012, alternative all-photonic schemes without quantum memories were presented by Munro \textit{et al.} \cite{Munro12} and by Azuma \textit{et al.} \cite{Azuma12, Azuma15} based on multi-photon cluster-states, loss-tolerant measurements and local high-speed active feedforward controls. Recently, Ewert \textit{et al.} achieved similar results using only \textit{locally}-prepared Bell states, instead of more demanding cluster states \textit{non-locally} entangled between nodes of a segment \cite{Ewert16}. One further scheme, feasible with current technology, was proposed in 2017 by Vinay \textit{et al.} \cite{Vinay17} based on double-heralded entanglement generation and brokered Bell-state measurements.

The third generation of QRs (III) employs QEC to deterministically correct errors from both propagation losses and gate errors \cite{Pant17, Azuma15, Fowler10, Munro12, Muralidharan14, Namiki16, Muralidharan17, Li13}. The idea is to iteratively transfer a block of $\sim 200$ qubits \cite{Munro12, Muralidharan14} from one node to the next in each lossy segment and to apply QEC to recover the encoding. Differently from the first two schemes (I, II), the latter is fully fault-tolerant and it involves one-way signalling between the segments, thus being much faster than the previous ones, in particular than type I. However, schemes II and III represent more demanding solutions from a purely technological perspective, mainly for the high control required to perform reliable error correction and, consequently, for the reduced maximal distance allowed between nodes.

\subsubsection{Blind quantum computing --}
\label{BQC}

In the previous sections we reviewed the physical resources for the implementation of a linear-optical quantum network. Entanglement plays a fundamental role for quantum communication, while various quantum repeaters have been developed with increasing performances, paving the way for an effective management of quantum information in intra-city and global networks. However, building a network with hundreds of nodes and synchronized single-photon sources seems still a demanding target for the near-term future. In this section we discuss quantum computation in an optical quantum network, where a client resorts to a central server with more advanced quantum technology to perform computation in a manner as safe as possible: the server should obtain no information on client's inputs, algorithms and outputs \cite{Fitzsimons17review}.

The first model of blind quantum computation (BQC) was proposed by Childs in 2001 \cite{Childs05}, where Alice, who cannot access a quantum computer, asks Bob to help her perform quantum computation without telling him her input, output and task. Childs described a protocol where Alice succeeds in this goal, with the possibility of even detecting whether Bob is honest or introducing errors. His solution was later improved by Arrighi and Salvail in 2003 \cite{Arrighi06}, with a blind protocol for the class of functions admitting an efficient generation of random input-output pairs like factoring. However, these schemes still allocate high resources to the client, who is supposed to have access to quantum memories and SWAP quantum gates. A step ahead was achieved in 2009 by Broadbent, Fitzsimons, and Kashefi with an interactive one-way BQC model (BFK) employing measurement-based quantum computing, where Alice only needs to produce single-qubit states with no need for quantum memories \cite{Broadbent09} and where she is capable to detect malicious errors (see Section \ref{sec:validation}).
From 2009 several other BQC protocols have been proposed, which may be classified according to the number of servers employed \cite{Sheng16}: \textit{single-server} BQC protocols \cite{Childs05, Arrighi06, Barz12, Dunjko12cv, Morimae12, Giovannetti13, Sueki13, Morimae13, Morimae14, Fisher14, Morimae15, Chien15, Gheorghiu15, Takeuchi16, Fitzsimons17, Mantri13, Mantri17}, where Alice is capable to generate quantum states and distribute them to Bob, \textit{two-server} \cite{Broadbent09, Sheng16, Morimae13server2} and \textit{three-server} \cite{Li14}, where Alice can get rid of quantum capabilities. Results have been reported also for a purely classical user \cite{Mantri17}, though some limitations have been recently opposed by Aaronson \textit{et al.} \cite{Aaronson17} building upon complexity considerations analogous to those for \textsc{BosonSampling} (see Section \ref{sec:BosonSampling}).
In measurement-only BQC, the idea is for Alice to secretly hide \textit{trap} qubits in her state. If  Alice discovers an unwanted change of a trap she can conclude that Bob is not honest and quit the task. The probability that Alice accepts a malicious Bob can then be made exponentially small using quantum error correcting codes \cite{Morimae14, Fitzsimons17}.

The feasibility of BQC has already been proved by various photonic experiments. The first demonstration was reported in 2011 by Barz \textit{et al.} \cite{Barz12}, where photon states were sent by the client to the quantum server, which produced four-photon blind cluster states to perform Grover search and Deutsch algorithms \cite{Nielsen_Chuang}. In 2013, Barz \textit{et al.} \cite{Barz13} proposed and demonstrated experimentally a technique to test whether a quantum computer is really quantum and whether it provides correct outputs (see Section \ref{sec:validation}). In the same year, Fisher \textit{et al.} \cite{Fisher14} demonstrated arbitrary computations on encrypted qubits, only requiring the client to prepare single qubits with limited classical communication. Finally, in 2016 Greganti \textit{et al.} \cite{Greganti16} reported photonic BQC where the client sends four-qubit states to the server to implement arbitrary two-qubit entangling gates.

The above demonstrations represent significant achievements for BQC, showing that integration of these algorithms on linear-optical networks are in principle feasible using single photons. We mention that also continuous variables schemes have been proposed \cite{Morimae12cv, Dunjko16}, while numerous works have analyzed their feasibility with respect to efficiency \cite{Mantri13, Giovannetti13, PerezDelgado15 } and resilience to errors \cite{Morimae12, Sueki13, Chien15} and to noisy implementations (see also Section \ref{sec:decoh.free}) \cite{Sheng16,Takeuchi16}.  However, it was observed that hybrid systems with matter qubits for carrying out quantum computation and single photons for quantum communication could offer an even more advantageous solution in larger-size distributed protocols \cite{Fitzsimons17review}.

\subsubsection{Photonic quantum key distribution --}
\label{sec:QKD}

Quantum key distribution (QKD) is a technique to generate a shared random secret key between two parties. The advantage of QKD is the possibility to detect possible attacks from a malicious third party, since any measurement carried out by the latter influences the shared system. Once a secret key is shared, the two parties can start standard classical communication.

\begin{table}[hb!]
	\renewcommand*{\arraystretch}{1.2}   
	\centering
	\footnotesize
	\caption{\label{QKDtable} }
	Several experimental demonstrations of quantum communication have been reported in the last two years, from specific applications to long-distance communication.
	\begin{tabular*}{\linewidth}{l@{\extracolsep{\fill}}ccc}
		\br
		Year	& Ref. 				&  Focus for QKD				& Distance (km) \\
		\mr
		2016$\:$	& \cite{Schiavon16} 		&  Renes2004 protocol 		& 0 			\\
			 	& \cite{Autebert16qkd} 	&  BBM92 protocol 			& 50 			\\
			 	& \cite{Sun16} 			&  WDM GPOM   			& 24	 		\\
				& \cite{Collins16} 		&  QDS   					& 90 			\\
				& \cite{Tang16} 		&  MDI DS					& 36 			\\
			 	& \cite{Yin16} 			&  MDI DS					& 404 		\\
			 	& \cite{Sun16decoy} 		&  Phase-encoded passive-DS   	& 10 			\\
			 	& \cite{Dynes16} 		&  DP-QPSK DS 				& 100    	 	\\
			 	& \cite{Lee16} 			&  HD DS DO  				& 43         		\\
			 	& \cite{Nape16} 		&  HD   					& 0 	    	  	 \\
			 	& \cite{Canas16} 		&  HD MCF DS   				& 0.3 		\\
			 	& \cite{Dynes16mcf} 		&  MCF   					& 53 			\\
			 	& \cite{Liao16} 			&  Free-space in daylight DS 	& 53     	  	 \\
		2017$\:$	& \cite{Zhang17QSDC} 	&  QSDC    				& 0        		\\
				& \cite{Sit17} 			&  Free-space HD   			& 0.3        		\\
				& \cite{Sibson17} 		&  DS on tunable circuit  		& 20 		 	\\
				& \cite{Ding17} 			&  HD MCF DS 				& 20 	       		\\
			 	& \cite{Sun17} 			&  Entanglement swapping		& 12   	  	 \\
			 	& \cite{Wang17}		&  WDM QAM DS   			& 80 	    		 \\
				& \cite{Collins17} 		&  QDS   					& 90/134	$^{(a)}$	\\
			 	& \cite{Roberts17} 		&  MDI DS QDS   			& 50 	     		  \\
			 	& \cite{Yin17} 			&  MDI DS QDS   			& 55     		  \\
			 	& \cite{Frohlich17} 		&  DS BB84				&  200/240 $^{(b)}$   \\
			 	& \cite{Kiktenko17} 		&  Polarization/phase-encoding  	& 45 	 		\\
			 	& \cite{Pugh17} 		&  Free-space  				& 10     	 	\\
			 	& \cite{Yin17_1200} 		&  Satellite  				& 1200 		    \\
			 	& \cite{Liao17} 			&  Satellite DS				& 700 	 		\\
			 	& \cite{Takenaka17} 		&  Satellite  				& 1000     	 	\\
			 	& \cite{Liao17satellite-to-ground}	&  Satellite DS 		& 1200 		    \\
			 	& \cite{Ren17}			&  Satellite teleportation 		& 1400 		    \\			 	
		2018$\:$	& \cite{Liao18} 			&  Satellite DS   				& 7600     		\\
		\br
	\end{tabular*}\\
	WDM: wavelength-division-multiplexing;\\GPOM: gigabit-capable passive optical network;\\MCF: multicore fiber; QDS: quantum digital signature;\\MDI: measurement-device-independent; DS: decoy-state; DP-QPSK: dual polarisation quadrature phase shift keying; QAM: quadrature amplitude modulation; HD: high-dimension;\\DO: dispersive optics;  QSDC: quantum secure direct communication.  $^{a}$: 134 km is simulated with additional optical attenuation. $^{b}$: with/without multiplexing.
\end{table}
\normalsize

\paragraph{Technological requirements and security.} The security of a QKD system is measured with the distance $\epsilon$ between the corresponding outcomes probability distribution and the ideal one with a perfect key, typically $\sim10^{-10}$. This difference can be made in principle arbitrarily small, at the price of requiring more stringent hardware performances and by applying suitable privacy amplification \cite{Nielsen_Chuang}; however, one should bear in mind that this threshold must include all subroutines involved in a QKD system, so that only the overall \textit{composable security}  $\tilde{\epsilon} =\sum \epsilon_k$ is relevant. There are still many limitations to overcome to develop QKD systems \cite{Diamanti16}, the main being (i) the transmission rate and range, much lower than in classical communications, (ii) the high cost to produce and maintain the hardware, (iii) the need for authentication and integrity and (iv) that new classical algorithms can be designed to be immune to quantum agents. First QKD prototypes have also been hacked \cite{Lo14,Moskovich15}, though techniques exist to correct gate errors with linear optics \cite{Mazurek14, Kalamidas05, Yamamoto05, Li07}.

QKD protocols are capable to establish secure communication channels, but no protection is warranted to the single-photon sources and detectors employed in the system. Alice's source represents probably the safest stage, since it can be secured by optical isolators and verified in situ \cite{Gottesman04}, thus most quantum hacking algorithms currently exploit the receiver's detectors (e.g. detection efficiency and dead time), even though a complete list of weak points would involve also attacks based on channel/device calibration or photon wavelength \cite{Lo14}.
In principle, countermeasures could be found any time a loophole is discovered just like for classical cryptosystems. However, this is naturally not desirable for an approach that aims to build inherently secure protocols of communication. One approach to counter hacking attacks is \textit{device-independent} (DI-) QKD \cite{Masanes11, Reichardt13} where Alice's and Bob's devices are considered as black boxes and the security depends on the violation of a Bell inequality to confirm quantum correlations\cite{Hensen15}. However, hardware technological challenges still make DI-QKD impractical for current state-of-the-art.
A second approach based on time-reversed QKD is \textit{measurement-device-independent} (MDI-) QKD \cite{Braunstein12, Lo12}, which allows Alice and Bob to perform QKD protocols even in the case of untrusted devices \cite{Inamori02}, -reasonably- assuming trusted sources. One advantage of MDI-QKD is its feasibility with state-of-the-art technology, with key rates orders of magnitude higher than in DI-QKD.

\paragraph{Implementation of QKD.} Currently the most widely adopted protocols for QKD are BB84 \cite{Bennett84} and E91 \cite{Ekert91}, as well as \textit{quantum secret sharing} \cite{Hillery99} and \textit{third-man quantum cryptography} \cite{Zukowski98}, though numerous other schemes have been developed \cite{Nielsen_Chuang, Bennett92, Hatakeyama17} based on discrete variables, continuous variable or distributed phase reference coding. Single photons offer the most promising platform to encode information that can be secretly shared thanks to entanglement. In particular, both quantum secret sharing and third-man quantum cryptography rely on three-particle polarization entangled states $\ket{\psi} \propto \ket{000}+\ket{111} $ known as GHZ \cite{Zukowski98}. Other schemes based on attenuated lasers are also of importance, for instance for differential phase shift \cite{Hatakeyama17, Collins16}, coherent one-way (COW) \cite{Gisin04} and decoy-state protocols \cite{SchmittManderbach07, Sun17, Canas16, Sun16decoy}. We refer the reader to recent comprehensive reviews \cite{Lo14, Diamanti16, Moskovich15} on the topic for a thorough overview of the extensive field. For a list of the most recent implementations see instead Table \ref{QKDtable}.

Information can be transferred between distant locations either via free-space ($\lambda\sim800$ nm) or via optical fibers ($\lambda\sim 1310$ nm or $\lambda\sim 1550$ nm) using any of the degrees of freedom overviewed in Section \ref{sec:encoding}. Polarization or orbital angular momentum are indeed more fit for free-space communication, since birefringence and mechanical instabilities in fibers can affect them heavily, thus making time-bin or frequency encoding more suitable. Protocols based on entangled photon pairs could achieve the longest distances tolerating higher losses up to $\sim70$ dB, as well as requiring higher technological requirements, while distributed-phase-reference QKD is a more promising solution for shorter distances  ($\sim100$ km)  \cite{Hatakeyama17, Gisin04}.\\
In the context of in-fiber QKD, a new approach that was recently introduced to achieve higher key rates is offered by wavelength-division multiplexing (WDM), where two quantum signals are transferred simultaneously on the same optical fibers. Quantum information can be exchanged even with strong classical signals thanks to dense wavelength multiplexing on the same fiber, showing that the integration of QKD protocols on existing telecom networks can be a viable path.  In 2009, Chapuran \textit{et al.} demonstrated WDM-QKD in a reconfigurable network with single photons at 1310 nm and classical channels at 1550 nm \cite{Chapuran09}, while Peters \textit{et al.} reported WDM-QKD at 1550 nm for both quantum and strong classical channels  \cite{Peters09}. More recently, in 2014 Patel and coworkers set a new record by transferring bidirectional 10 Gb/s classical channels with a key rate of 2.38 Mbps and fiber distances up to 70 km \cite{Patel14}. One of the main sources of noise in these experiments was due to spontaneous anti-Stokes Raman scattering \cite{Peters09, Chapuran09}, whose influence was shown to be mitigated by selecting an optimal wavelength, allowing for unprecedented terabit classical data transmission up to 80 km \cite{Sun16, Wang17}.

While in-fiber QKD could achieve high transmission rates for short distances, by exploiting pre-existing telecom infrastuctures, free-space QKD represents an alternative promising way for future QKD networks on a global scale. First demonstrations of ground-based quantum communication with single photons were reported with progressively higher records in distance and key rate \cite{Aspelmeyer03, Peng05, Resch05, Ursin07, SchmittManderbach07, Jin10, Yin12, Wang12oam, Heindel12, Rau14, Vallone14, Steinlechner17}. Yet, one further promising approach would be to employ satellites as nodes of the network, an ambitious project that could allow to cover the whole globe. Preliminary tests in this direction were carried out independently in 2013 by Wang \textit{et al.}, simulating three experiments of QKD with decoy states \cite{Canas16, SchmittManderbach07, Sun16decoy} with a setup operating on moving (ground) and floating (hot-air balloon) platforms with high-loss channels \cite{Wang13satellite}, and by Nauerth \textit{et al.}, demonstrating an instance of BB84 between a high-speed airplane and a ground station for a distance of 20 km \cite{Nauerth13}. Later on, in 2015 and 2016, two papers reported the transmission of  single photons using satellite corner cube retroreflectors as quantum transmitters in orbit \cite{Vallone15, Dequal16}. In 2017 six further achievements were reported: ground-to-aircraft \cite{Pugh17} and satellite-to-ground \cite{Liao17, Takenaka17, Liao17satellite-to-ground} QKD, satellite-based entanglement distribution over 1200 kilometers \cite{Yin17_1200} and ground-to-satellite teleportation of single-photon qubits for a distance up to 1400 km \cite{Ren17}. Finally, in 2018 Liao \textit{et al.} successfully performed decoy-state QKD between a low-Earth-orbit satellite and multiple ground stations separated by 7600 km on Earth \cite{Liao18}, providing landmark steps forward toward the realization of future long-distance quantum networks.


\section{Photonic quantum simulation}
\label{sec:simulation}

In this chapter we will review the field of photonic quantum simulation, starting with single-photon dynamics in quantum walks (Section \ref{sec:QW}) and extending it to multiphoton evolution (Section \ref{sec:BosonSampling}). In Section \ref{sec:validation} we will review some of the most recent results on the problem of verification, whose increasing relevance goes in parallel to the developments in quantum simulation. Finally, in Section \ref{qChemistry} we describe the first applications in quantum chemistry and condensed matter. For a comprehensive review of quantum simulation in various fields and physical platforms we refer the reader to Refs. \cite{Buluta09, Georgescu14}.

\subsection{Photonic quantum walks}
\label{sec:QW}

\begin{figure*}[t!h]
\centering
\includegraphics[trim={0 0 0 0},clip, width=\textwidth]{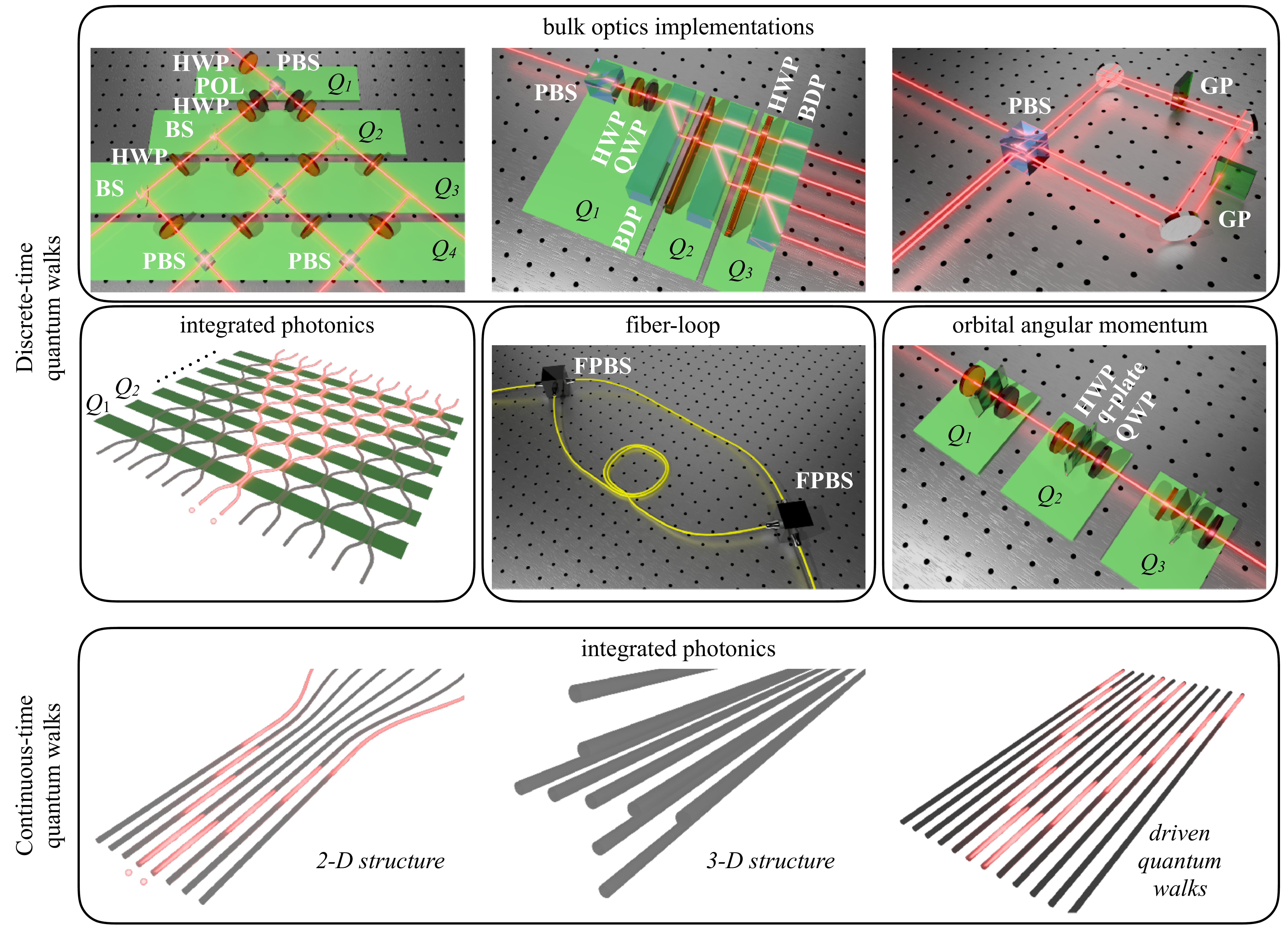}
\caption{\footnotesize Quantum walks are classified in two types: \textit{discrete-time} and \textit{continuous-time}. While the former can count on various photonic implementations, the latter has been enabled by integrated waveguide lattices. In each subfigure, $Q_i$ indicates the different steps of the quantum walks. Legend - QWP: quarter-wave plate, HWP: half-wave plate, PBS: polarizing beam splitter, BS: beam splitter, GP: glass plate, POL: polarizer, FPBS: fiber polarizing beam splitter.}
\label{fig:FigureQuantumWalk}
\end{figure*}

Quantum walks (QW), the extension of classical random walks to a quantum framework, have gained an increasing role in modelling single- and multi-particle evolutions in several scenarios \cite{Grafe16} thanks to their platform-independent formulation. Importantly, the universality of the linear-optical platform for quantum computation \cite{KLM} has further prompted several experimental tests of QWs with single photons.
Two classes of QWs exist (see Fig.\ref{fig:FigureQuantumWalk}): \textit{discrete-time} (DTQW), where the evolution is split in discrete steps and random events suddenly influence the dynamics, and \textit{continuous-time} (CTQW), where the evolution on a given lattice is described by a time-independent Hamiltonian and by the adjacency matrix of the corresponding graph. In this section we will focus on the experimental implementation of photonic QWs, reviewing the latest achievements in this area with some links to the theoretical background. The reader interested in the theory of quantum walks may refer to Ref. \cite{Venegas-Andraca12} for a comprehensive collection.

\paragraph{Discrete-time quantum walks.}

DTQWs can be described by single photons evolving through a network of beam splitters. In this pictorial representation, each time a photon enters a beam splitter its wave function is split in two parts proceeding on separate optical modes. In terms of creation operators acting on the two modes, the evolution can be ruled for instance by the \textit{Hadamard coin}

\begin{equation}
	\left(
	\begin{array}{c}
		a^{\dagger}_1\\
		a^{\dagger}_2
	\end{array} \right)_{in}
	\longrightarrow \quad
	\frac{1}{\sqrt{2}} \left(
	\begin{array}{cc}
		1& i \\
		i & 1
	\end{array} \right)
	\left(
	\begin{array}{c}
		a^{\dagger}_1\\
		a^{\dagger}_2
	\end{array} \right)_{out}
\end{equation}

\noindent We stress that also other approaches can accomplish the same task \cite{Cardano15, Cardano16}: we only need a process \^{Q} that, given a chosen encoding, operates the transformation $Q \ket{1} \longrightarrow \alpha \ket{0} + \beta \ket{1} $. Choosing \^{Q} as a building block for the QW, after $n$ steps the walker will be in the superposition $ \sum_{j}^{n} c_j\, a^{\dagger}_j \ket{0} $, where the amplitudes $c_j$ can be balanced, symmetric or asymmetric over the optical modes according to the symmetries of the walker \cite{Sansoni12, Carolan14} and to the evolution \cite{Schreiber11, Crespi13anderson}.

First implementations of DTQWs were reported in the first years of this century, motivated by the connection found between QWs and quantum computation \cite{Childs03, Childs09} for which, just at that time, Knill, Laflamme and Milburn were showing that linear-optical platforms are universal \cite{KLM}. The very first demonstrations were carried out on bulk optics, confirming the theoretical predictions for single photons dynamics \cite{Do05, Broome10} and wave packet interference \cite{SoutoRibeiro08}. However, bulk implementations unavoidably suffer from mechanical instabilities and issues of alignment critical for larger-size experiments. A first solution was found in 2010 with novel fiber-loop designs \cite{Schreiber10,Schreiber11, Regensburger11, Schreiber12}, which provided a more stable approach suitable for further improvements in the direction of scalable schemes. Thanks to this new approach it was possible to investigate QWs with a larger number of steps, yielding a richer landscape of photon interference. A relevant step forward was given by the introduction of integrated photonic circuits (see Section \ref{sec:circuits}), which provided furter advantages in stability, alignment and compactness \cite{Sansoni12, Crespi13anderson, Carolan14, Grafe14, Harris17}. Moreover, integrated circuits are fit for realizing reconfigurable photonic architectures, allowing to dynamically tune the evolution and, thus, to explore more complex scenarios within the same quantum walk \cite{Carolan15, Flamini15, Harris17}. Relevant tests on the foundations of quantum mechanics carried out on integrated circuits include, for instance, bosonic-fermionic evolutions \cite{Sansoni12}, indistinguishable-distinguishable bosons \cite{Carolan14} and quantum trasport phenomena \cite{Harris17} such as wavefunction localization in disordered media \cite{Crespi13anderson}. Very recently, in 2015 an approach for DTQWs was reported based on the orbital angular momentum of light, so that the evolution takes place in the OAM degree of freedom without the need for an interferometer \cite{Cardano15, Cardano16}. This new approach provides indeed a flexible solution that may enable scalable investigations. Finally, in 2016 Boutari \textit{et al.} proposed and demonstrated experimentally a time-bin-encoded scheme for QWs, based on a network of optical ring cavities, which simultaneously achieves low losses, high fidelity, reconfigurability. Such scheme can be further generalized to multidimensional lattices \cite{Boutari16}.

\begin{table*}[h!]
	\renewcommand*{\arraystretch}{1.2}   
	\centering
	\footnotesize
	\caption{\label{QWtableDT} Photonic discrete-time quantum walks (QW).}
	\begin{tabular*}{\textwidth}{c@{\extracolsep{\fill}}ccccc}
		\br
		Year	 & Ref. 				&  Platform			& Focus 						& Photons 			& Structure	\\
		\mr
		2005	 & \cite{Do05} 			&  Bulk optics			& Quantum quincunx				& 1 				& 8 modes		 \\
		2008	 & \cite{SoutoRibeiro08} 	&  Bulk optics			& Wave packet reshaping			& 1 				& 1 step		 \\
		2010	 & \cite{Schreiber10}		&  Fiber loop			& Robust implementation			& 2 				& 5 steps		 \\
			 & \cite{Broome10}		&  Bulk optics			& QW with tunable decoherence		& 1 				& 6 steps 		\\
		2011	 & \cite{Regensburger11}	&  Fiber loop			& Features of the dynamics			& Attenuated laser 	& 70 steps		 \\
			 & \cite{Schreiber11} 		&  Fiber loop			& Decoherence and disorder		& 1				& 28 steps		 \\
		2012	 & \cite{Kitagawa12} 		&  Bulk optics			& Topological phases				& 1				& 7 steps 		\\
			 & \cite{Sansoni12} 		&  Integrated photonics	& Bosonic-fermionic evolution 		& 2 				& 8 modes		 \\
			 & \cite{Schreiber12} 		&  Fiber loop			& Entanglement on 2D QW		  	& Attenuated laser	& 12 steps 	\\
		2013	 & \cite{Crespi13anderson} &  Integrated photonics	& Anderson localization			& 2  				& 16 modes	 \\
			 & \cite{Jeong13} 		&  Fiber loop 			& Delayed-choice 2D QW, Grover QW	& 1  				& 4 steps 		\\
		2014	 & \cite{Grafe14} 		&  Integrated photonics 	& High-order single-photon W-states  	& 1				& 2,4,5,8,16 modes \\	
		2015	 & \cite{Cardano15} 		&  Orbital angular momentum	& Wave packet dynamics 			& 2			& 5 steps \\
		2016	 & \cite{Boutari16} 		&  Optical ring cavities	& Low-loss tunable QW  			& Attenuated laser	& 62 steps \\
			 & \cite{Cardano16} 		&  Orbital angular momentum 	& Topological quantum transitions  	& 1			& 6 steps \\
		2017  & \cite{Cardano17} 		&  Orbital angular momentum 	& Zak phases, topological invariants  	& Attenuated laser	& 7 steps \\		
			 & \cite{Pitsios17} 		&  Integrated photonics 	& Entanglement after spin chain quench  	& 2		& 5 modes \\
			 & \cite{Harris17} 		&  Integrated photonics	& Quantum transport 				& Attenuated laser	& 26 modes \\
		\br
	\end{tabular*}\\
\end{table*}
\normalsize

\paragraph{Continuous-time quantum walks.}

\begin{table*}[th!]
	\renewcommand*{\arraystretch}{1.2}   
	\centering
	\footnotesize
	\caption{\label{QWtableCT} Photonic continuous-time quantum walks implemented with integrated circuits.}
	\begin{tabular*}{\textwidth}{c@{\extracolsep{\fill}}ccccc}
		\br
		Year 		& Ref. 			& Technology 				& Focus 						& Photons 			& Waveguides \\
		\mr
		2007	 	& \cite{Schwartz07} 	&  Optical induction$^{(a)}$ 	& Transport, Anderson localization	& Attenuated laser 	& n.a.	\\
		2008	 	& \cite{Perets08} 	&  SoI					& Evolution on quantum walks  		& Attenuated laser & $\sim100$	\\
		2010	 	& \cite{Peruzzo10} 	&  $\textup{SiO}_x\textup{N}_y$ & Quantum correlations  			& 2				& 21		\\
		2011		& \cite{Owens11}	&  FLW					& Evolution in an elliptic waveguide array	& 2 			& 6		\\
				& \cite{Martin11}	&  FLW					& Anderson localization			& Attenuated laser	& 101	\\
		2012	 	& \cite{Lahini12}	&  n.a.			 		& Quantum correlations	 		& Attenuated laser	& 29		  \\
				& \cite{Crespi12}	&  FLW					& Quantum Rabi model			& Attenuated laser 	& 15 		\\
		2013	 	& \cite{Rechtsman13} &  FLW 					& Floquet topological insulator  		& Attenuated laser 	& 49/308 $^{(b)}$	\\
				& \cite{Spagnolo13} 	&  FLW 					& Bosonic coalescence			& 3 				& 3		\\
				& \cite{Matthews13}	&  $\textup{SiO}_x\textup{N}_y$	& Fermionic statistics				& 2 				& 10		\\
		2014		& \cite{Solntsev14}	&  $\textup{LiNbO}_3$		& Tunably-entangled biphoton states 	& 2 				& 21 		\\
			  	& \cite{Carolan14} 	&  $\textup{SiO}_x\textup{N}_y$ & Bosonic clouding				& 3,4,5   			& 21 		\\
				& \cite{Poulios14}	&  FLW					& Quantum correlations			& 2 				& 9		\\
		2015	 	& \cite{Keil15} 		&  FLW 					& Majorana dynamics  			& Attenuated laser	& 26		\\
				& \cite{Crespi15fano}	&  FLW					& Quantum decay and Fano interference	& 2 			& 27 	\\
			 	& \cite{Lebugle15} 	&  FLW 					& N00N state Bloch oscillations		& 2 				& 16		\\
		2016	 	& \cite{Biggerstaff16} &  FLW 					& Quantum transport				& Attenuated laser	& 4+30	\\
			 	& \cite{Caruso16} 	&  FLW 					& Quantum transport				& Attenuated laser 	& 18+62	\\
			 	& \cite{Weimann16} 	&  FLW 					& Discrete fractional Fourier transform	& 2 				& 8		\\
		\br
	\end{tabular*}\\
	FLW: femtosecond laser written. SoI: silicon on insulator. n.a.: not available.\\$^{(a)}$:  in $\textup{Sr}_{0.6}\textup{Ba}_{0.4}\textup{Nb}_{2}\textup{O}_{6}$; (b): for two honeycomb photonic lattices. 

\end{table*}
\normalsize

CTQWs represent another platform for investigating statistical features of the evolution of bosonic states \cite{Schwartz07, Perets08, Owens11, Biggerstaff16, Caruso16} and simulating complex processes \cite{Crespi12, Spagnolo13, Rechtsman13, Matthews13, Keil15, Crespi15fano, Lebugle15}. CTQWs can be implemented with arrays of $N$ evanescently-coupled waveguides tailored to produce a time-dependent or time-independent lattice Hamiltonian  \cite{Bromberg09}

\begin{equation}
\check{H} = \hbar \sum_{i=1}^N \left(   \beta_i \, a_i^\dagger a_i + \sum_{j=1}^N \kappa_{i,j} \, a_j^\dagger a_i  \right)
\end{equation}

\noindent where $\beta_i$ is the propagation constant for the $i$th waveguide and $\kappa_{i,j}$ is the coupling between waveguides ($i$,$j$), with the usual restriction to $j=i\pm 1$.\\For an analogy with DTQWs, where the free evolution is decomposed in discrete steps, now waveguides are arranged in 1D or 2D structures in such a way that light can continuously jump between neighbor sites.
The evolution of the $k$th creation operator is given by \cite{Bromberg09}

\begin{equation}
\cases{
	a^{\dagger}_k(z) = e^{i \beta z} \sum_l  U_{k,l}(z) \;a^{\dagger}_l (z_0) \\ U_{k,l}(z) = (e^{i \kappa z})_{k,l}
	}
\end{equation}

\noindent where $z$ is the position along the propagation direction. Waveguide arrays have been implemented so far on integrated photonic circuits, inscribed using one of the fabrication techniques described in Section \ref{sec:circuits}.

\paragraph{Quantum transport.}

Quantum transport phenomena concern the spatial evolution of the wave function during a QW. This process can occur in two regimes: in \textit{ordered} and in \textit{disordered} lattices. A photon propagating through a static-ordered lattice shows a \textit{ballistic} spread for a distance proportional to the evolution time $T$, due to the interference of the wave packet amplitudes across the CTQW. For disordered systems a further distinction can be made: \textit{dynamic disorders} are associated to \textit{diffusive} transports, where interference no longer affects the evolution and a classical-like spreading is found for a distance proportional to $\sqrt{T}$, while \textit{static disorders} lead to a complete halting in the process, a phenomenon known as Anderson localization \cite{Segev13, Abouraddy12} where the wave packet localizes on the initial sites of the lattice \cite{Schwartz07, Lahini08, Martin11, Crespi13anderson}. Several analyses show that a suitable combination of quantum coherence and environmental noise can provide an effective enhancement in quantum transport \cite{Rebentrost09, Caruso09, Caruso11, Viciani15, Novo16, Biggerstaff16, Harris17}. In this context, reconfigurable photonic circuits can provide a useful platform to characterize it, allowing to study the continuous transitions between different regimes \cite{Harris17}. Finally, we mention that lattices subject to external gradient forces exhibit a characteristic periodic oscillation between spreading and localization of the wave packet, the so-called \textit{Bloch oscillations}, which have been observed experimentally in photonic platforms \cite{Morandotti99, Pertsch99, Dreisow09, Regensburger11, Lebugle15}.

\paragraph{Multi-photon quantum walks.}

So far we have reviewed studies that focus on single-particle evolutions in discrete- or continuous-time QWs. However, the full landscape of phenomena gets enriched if we consider photonic states with $n=2$ photons \cite{Crespi13anderson, Sansoni12, Schreiber10, Cardano15, Owens11, Spagnolo13, Matthews13, Solntsev14, Peruzzo10, Poulios14, Weimann16, Carolan14, Crespi15fano, Lebugle15}, thanks to the emergence of more complex interference patterns \cite{Weimann16, Carolan14}, highlighting for instance two-photon quantum correlations \cite{Peruzzo10, Poulios14} with bosonic-fermionic transitions \cite{Sansoni12, Matthews13}, and enabling the possibility of simulating relevant physical processes \cite{Crespi15fano, Lebugle15}.
Finally, the possibility of further increasing the number of injected photons to $n=3$ \cite{Metcalf13, Spagnolo13} or $n>3$ photons is believed to disclose even larger potentialities, which we briefly discuss in the next section with the \textsc{BosonSampling} problem.

\subsection{\textsc{BosonSampling}}
\label{sec:BosonSampling}

In the previous section we reviewed the recent achievements in photonic quantum walks, which gained increasing attention after the discovery that multi-particle quantum walks with interactions allow for universal quantum computation \cite{KLM, Childs13}. A milestone result in the intertwined developments of quantum walks and quantum computation is represented by the proof, by Aaronson and Arkhipov \cite{AA} in 2010, that $n$-photon states evolving in a discrete-time quantum walk can provide a first experimental evidence of a superior quantum computational power \cite{Gard15, Harrow17}. 

\textsc{BosonSampling} consists in sampling from the output distribution of an interferometer implementing a Haar-random $m$-mode transformation $U$ on $n$ indistinguishable bosons. The computational complexity of the problem is rooted in the hardness \cite{Valiant79} of evaluating the permanent in the scattering amplitudes \cite{Scheel08}
  
\begin{equation}
	\bra {t_1...t_m} \hat U \ket{s_1...s_m} = \left( \prod_{i=1}^m s_i! \, t_i! \right)^{-\frac{1}{2}} \textup{Per\,} (U_{S,T})
\end{equation}

\noindent where the integer $s_k$ ($t_k$) is the occupation number for the input (output) mode $k$ ($\sum s_k=\sum t_k=n$) and $U_{S,T}$ is the $n \times n$ matrix obtained by repeating $s_k$ ($t_k$) times the $k$th column (row) of $U$. Assuming two highly plausible conjectures \cite{AA}, Aaronson and Arkhipov showed that, should a classical polynomial-time algorithm exist capable to solve \textsc{BosonSampling}, it would imply the collapse of the polynomial hierarchy of complexity classes to the third level, a possibility of huge consequences widely believed to be unlikely.

Despite its simple formulation, requiring only single photons with no entanglement, no adaptive measurements and no ancillary qubits, the necessity to scale up the number of photons and modes represents a technological challenge. For an overview of the state of the art in the development of single-photon sources, integrated photonic interferometers and single-photon detectors, the reader can refer to Sections \ref{sec:sources}, \ref{sec:circuits} and \ref{sec:detectors}.
State-of-the-art classical simulation of \textsc{BosonSampling} depends on the number of photons as in Table \ref{table:BShardness}.

\begin{table}[h!] 
	\renewcommand*{\arraystretch}{1.25}                       
	\centering
	\caption{\label{table:BShardness}
		The largest classical simulation of \textsc{BosonSampling} (up to 30 photons) was reported in 2017 based on Metropolised independence sampling (MIS) \cite{Neville17}. The computational complexity for exact classical \textsc{BosonSampling} was given in the same year in Ref. \cite{Clifford18} by Clifford and Clifford (CC): state-of-the-art sampling time is equal, within a factor of 2, to computing the permanent of one single scattering matrix \cite{Wu16}.
	}
	\footnotesize                 
	\begin{center}
		\begin{tabular}{cccc}
			\mr
			Approach				& $n\sim$		& Hardware			& Classical technique \\
			\mr		
			Simulated				& 30			& Laptop				& MIS \cite{Neville17}	  \\ \hline
			\multirow{ 2}{*}{Exact} 	& 10			& Laptop				& Brute force	  \\
								& 50			& Tianhe-2 \cite{Wu16}	& CC \cite{Clifford18}  \\
			\mr
		\end{tabular}
	\end{center}
\end{table}

Soon after its introduction, in 2013 four experimental demonstrations with $n=3$ photons were reported \cite{Broome13, Crespi13, Spring13, Tillmann13}. Since then, several investigations have been performed to study the scalability in imperfect conditions, such as in the presence of losses  \cite{Aaronson15, Motes15, Garcia-Patron17, Oszmaniec18}, partial distinguishability \cite{Shchesnovich15, Tillmann15} and generic experimental errors \cite{Leverrier15, Motes15, Rohde12, Shchesnovich14}. Furthermore, the scalability of parametric down-conversion sources and the relevance of non-ideal detectors' efficiencies have been addressed in Ref.\cite{Motes13} and investigated experimentally in 2015 with the \textit{Scattershot} implementation \cite{Lund14, Bentivegna15}.

\begin{table*}[h!]
	\renewcommand*{\arraystretch}{1.15}   
	\centering
	\footnotesize
	\caption{\label{BStable} Experimental demonstrations of photonic \textsc{BosonSampling}.}
	\begin{tabular*}{\textwidth}{c@{\extracolsep{\fill}}cccccc}
		\br
		Year & Ref. & Technology & Features & Photons & Modes &Validation\\
		\mr
		2013	 & \cite{Broome13} 		& in-fiber  		&						& 2, 3 				& 6		 	&    \\
			 & \cite{Crespi13} 		& FLW  		&  						& 3 					& 5 			&    \\
			 & \cite{Spring13} 		& SoS  		&  						& 3		 			& 6 			&   \\
			 & \cite{Tillmann13} 		& Si  		&  						& 3 					& 5 			&    \\
		2014	 & \cite{Spagnolo14} 		&  FLW 		& First validation 			& 3  					& 5, 7, 9, 13  	& U, D   \\
			 & \cite{Carolan14} 		& $\textup{SiO}_x\textup{N}_y$  	& Bosonic clouding & 3 			& 9 			& U, D   \\
		2015	 & \cite{Carolan15} 		& SoS  		& Full reconfigurability 		& 3		 			& 6 			& D, MF \\
			 & \cite{Tillmann15} 		& FLW  		& Full interference spectrum  	& 3 					& 5 			&    \\
			 & \cite{Bentivegna15} 	& FLW  		& Scattershot \textsc{BosonSampling}	& 3 			& 9, 13 		& U, D \\
		2016	 & \cite{Crespi16} 		& FLW  		& 3D Fourier interferometer 	& 2 					& 4, 8 		& D, MF   \\
		2017	 & \cite{Loredo17} 		& in-fiber  		& High purity and brightness 	& 2, 3				& 6 			& U, D   \\
			 & \cite{He17} 			& Bulk (time-bin) & High purity and brightness 	& 3, 4 				& 6, 8 		& U, D, MF   \\
			 & \cite{Wang17bs} 		& Fused-quartz 	& High purity and brightness 	& 3, 4, 5 				& 9 			& U, D   \\
		\br
	\end{tabular*}\\
	U, D, MF: 'Uniform', 'Distinguishable' and 'Mean-Field' samplers.\\
	SoS: silica on silicon. FLW: femtosecond laser written. \\
	Some demonstations have reduced computational complexity due to bunched input states $\ket{2, 2}$ \cite{Spring13} and $\ket{3, 3}$ \cite{Carolan15}, in the case of standard \textsc{BosonSampling} \cite{AA}, or due to partial coverage of the full Hilbert space in the Scattershot version \cite{Bentivegna15}.
\end{table*}
\normalsize

\begin{figure}[t!]
\centering
\includegraphics[trim={0 0 0 0},clip, width=0.5\textwidth]{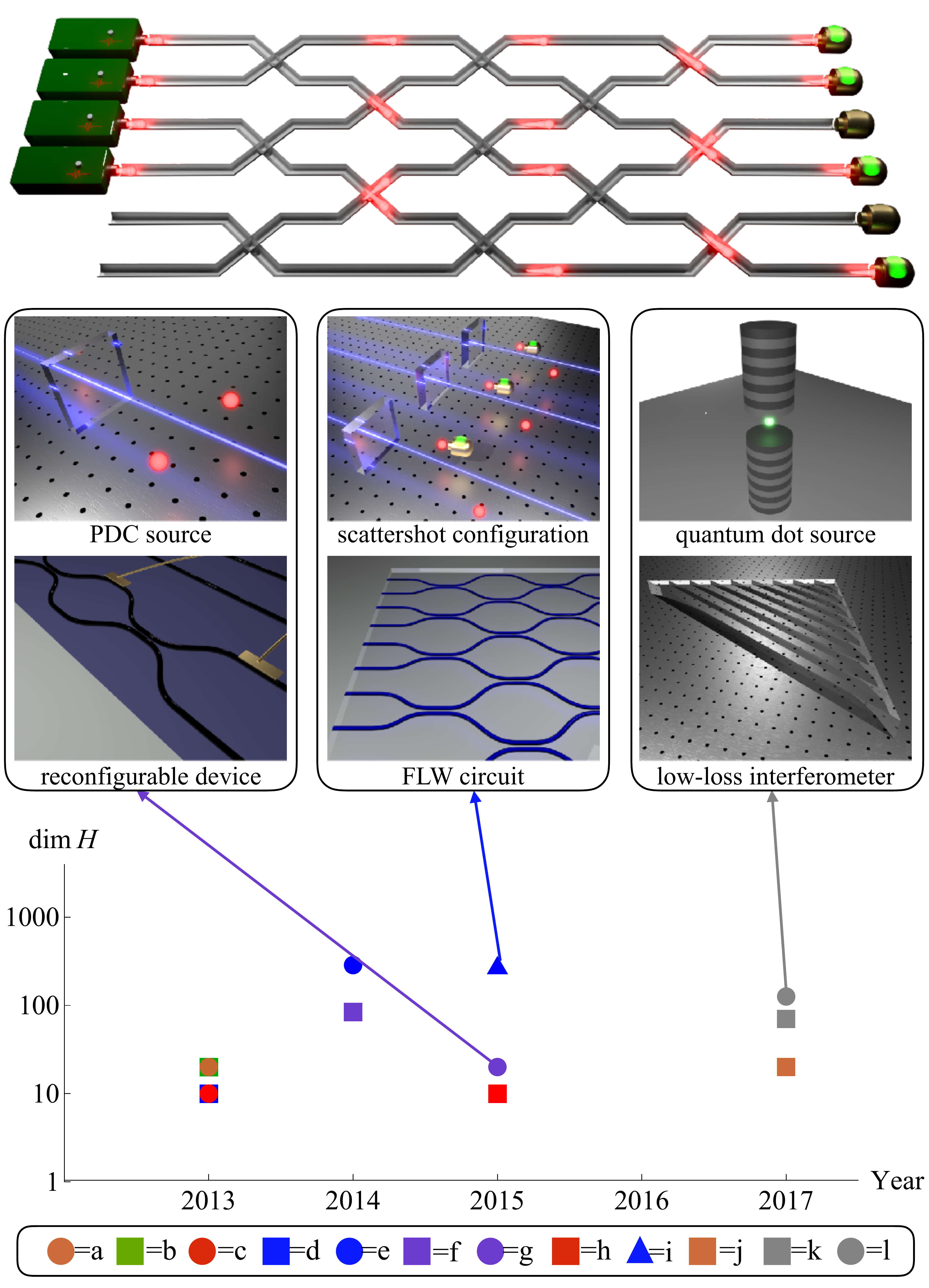}
\caption{\footnotesize Increase of the Hilbert-space dimensionality $\mathrm{dim} H$ in \textsc{BosonSampling} experimental implementations with single photons. Top inset: scheme of an integrated photonic device for \textsc{BosonSampling}, where a no-collision state is injected in the first optical modes and measured at the output after a Haar-random unitary evolution. Legend - PDC: parametric down-conversion, FLW: femtosecond laser writing. Recently the implementation of a lossy \textsc{BosonSampling} experiment has been reported \cite{Wang18}.\\Bottom legend: references to the related literature.\\ a: \cite{Broome13}. b: \cite{Spring13}. c: \cite{Tillmann13}. d: \cite{Crespi13}. e: \cite{Spagnolo14}. f: \cite{Carolan14}. g: \cite{Carolan15}. h: \cite{Tillmann15}. i: \cite{Bentivegna15}. j: \cite{Loredo17}. k: \cite{He17}. l: \cite{Wang17bs}}
\label{fig:FigureBosonSampling}
\end{figure}

Since 2015, a number of alternative schemes have been proposed: driven \textsc{BosonSampling}, where the input photons are generated within the interferometer \cite{Barkhofen17}, \textsc{BosonSampling} with microwave photons \cite{Peropadre16} or squeezed states \cite{Hamilton17, Kruse18}, one using Gaussian measurements and the symmetry of the evolution under time reversal \cite{Chakhmakhchyan17, Chabaud17}, and one time-bin loop-based \cite{Motes14timebin}. Furthermore, somehow in analogy with the link between quantum walks and quantum computation, recently \textsc{BosonSampling} was shown to be equivalent to short-time evolutions of $n$ excitations in a XY model of 2$n$ spins \cite{Peropadre16}. This feature has suggested multiport photonic interferometers as good candidates for the implementation of quantum simulators or even general-purpose quantum computers \cite{Peropadre16, GonzalezAlonso16bs, Huh15}. Nevertheless, in 2017 new classical algorithms have been proposed to solve \textsc{BosonSampling} for systems with dimensionality larger than that allowed by near-term technological advances \cite{Neville17, Clifford18, Renema17, Garcia-Patron17, Chakhmakhchyan17alg, Oszmaniec18}, thus increasing the challenge to achieve quantum supremacy, i.e. the regime with quantum advantage \cite{Gard15, Harrow17}. Experimental state of the art for photonic \textsc{BosonSampling} is shown in Table \ref{BStable} and in Fig. \ref{fig:FigureBosonSampling}.

\subsection{Verification of quantum simulation}
 \label{sec:validation}

The problem of verification, fundamental for both classical and quantum computation, is closely related to the issue of security in information processing. 
One good example is offered by \textsc{BosonSampling} \cite{AA}: a computational problem where a special-purpose quantum device is believed to outperform classical supercomputers. For similar problems that live outside the complexity class NP, it is in general not possible to verify the correctness of an outcome. Thus, in a scenario where untrusted parties can deviate calculations from their ideal flow, it is essential to develop techniques to ensure that robust computation is guaranteed to the maximum possible extent (fault-tolerance, i.e. amplification of the detection rate) or, equivalently, that malicious processes can be easily detected (e.g. using traps to detect errors) \cite{Broadbent09}. In this sense, the problem of verification in quantum computation is connected with the universal blind computation model described in Section \ref{BQC}. In the last few years, several protocols for verifiable quantum computations have been proposed \cite{Broadbent09, Reichardt13, Morimae14, Morimae14oneclean, Hajdusek15, Gheorghiu15, Fitzsimons15, Kapourniotis15, Kashefi15, Hayashi15, Morimae16verification, Gheorghiu17}, where the attention focused on schemes with either fewer experimental requirements or higher generality and security. We refer the reader to Ref.\cite{Fitzsimons17} for a comprehensive review on the problem of verification in blind quantum computing and to Refs. \cite{Kashefi15, Kapourniotis15} for a short overview of the differences between the various protocols.

We have shortly described in Section \ref{BQC} the first experimental demonstrations of verifiable blind quantum computing, performed using polarization qubits represented by multi-photon states generated via parametric down-conversion \cite{Barz12, Barz13, Fisher14, Greganti16}. 
Beyond these works, the problem of verification has drawn large attention also in quantum communication and quantum simulation. In the scope of secure quantum communication, for instance, where quantum networks are employed to perform distributed quantum computations over possibly untrusted nodes, it is necessary to verify the goodness of the shared multipartite entangled states, the key resource behind these protocols \cite{Pappa12, Grafe14}.
To this aim, in 2014 Bell \textit{et al.} reported the experimental demonstration of a graph-state quantum secret sharing protocol \cite{Bell14}, whose implementation was subsequently verified using the scheme in Ref.\cite{Pappa12}. In 2016, further improvements led McCutcheon \textit{et al.} to a verification of multipartite entanglement for photonic quantum networks, adopting two single-photon sources similar to Ref.\cite{Bell14} to produce three- and four-photon polarization encoded states \cite{McCutcheon16}.

We conclude this section discussing the role of verification in quantum simulation and, in particular, in \textsc{BosonSampling}. In the latter case, as for analogous experiments involving multi-photon interference in linear-optical multiport interferometers, it is crucial to check the indistinguishability of the input photons to ensure the correctness of the overall operation. The \textit{validation} of \textsc{BosonSampling} developed rapidly in the last few years just after the first experimental demonstrations, and its applications will potentially go beyond its original purpose. Since a full certification of \textsc{BosonSampling} is believed to be not possible \cite{Aaronson14}, all current protocols aim at ruling out the most plausible unwanted scenarios, namely experiments where devices are injected with distinguishable input photons and with mean-field states \cite{Tichy14}. Currently there exist several protocols to validate \textsc{BosonSampling}, most of which have already been successfully demonstrated experimentally \cite{Spagnolo14, He17, Wang17, Crespi16, Carolan15, Viggianiello17sys, Viggianiello17tvd, Bentivegna14, Bentivegna15, Carolan14, Agresti17, Wang18, Giordani18}. An efficient certification protocol for photonic state preparation was also introduced theoretically \cite{Aolita15}, allowing to discriminate relevant classes of Fock states and Gaussian/non-Gaussian pure states. We summarize them in Table \ref{tableValidation}, with an overview of the techniques adopted and the hypotheses which are designed for.

\begin{table}[b!]
        \renewcommand{\arraystretch}{1.2}
        \footnotesize                  
	\centering                      
	\begin{center}
		\begin{tabular*}{\linewidth}{c@{\extracolsep{\fill}}ccccc}
			\br
			 Test  & Ref.  & U & D & MF & O \\
			\mr
			\ RNE  	 & \cite{Aaronson14, Spagnolo14, He17, Wang18, Bentivegna15} & $\checkmark$	& 				& 				& 				 \ \\
			\ LR		 & \cite{Spagnolo14, Wang17bs, He17, Wang18, Bentivegna15} & $\checkmark$	& $\checkmark$		& 				& $\checkmark$ 	 \ \\
			\ ZTL		 & \cite{Tichy14, Dittel17, Dittel18, Crespi15, Crespi16, Carolan15, Viggianiello17sys, Viggianiello17tvd}  & 			& $\checkmark$		& $\checkmark$	 	&			 	\ \\
			\ Bayesian 	 & \cite{Bentivegna14, Wang17bs, He17, Viggianiello17tvd}& $\checkmark$	& $\checkmark$		& 				& $\checkmark$	  	 \ \\
			\ Bunching  & \cite{Carolan14, Shchesnovich16} & 			& $\checkmark$		& 	 	&			 \ \\
			\ CG		 & \cite{Wang16bubbles, Agresti17}  & $\checkmark$ & $\checkmark$		&				& $\checkmark$ 	\ \\
			\ $n=m$	 & \cite{Liu16} & 			& $\checkmark$		& $\checkmark$		& 				 \ \\
			\ Statistical & \cite{Walschaers16, Giordani18}  &	$\checkmark$		& $\checkmark$		& $\checkmark$		& $\checkmark$  	 \ \\
			\mr
		\end{tabular*}
	\end{center}
	\caption{Validation of multi-photon interference against uniform distribution (U), experiments with distinguishable photons (D), mean-field states (MF) or other hypotheses (O). See Table \ref{BStable} for a cross reference with \textsc{BosonSampling} experiments.\\
	RNE: row-norm estimator; LR: likelihood ratio; ZTL: zero-transmission law (suppression law); CG: Coarse-graining.  }
	\label{tableValidation}
\end{table}

\subsection{Photonic simulation in quantum chemistry and condensed matter}
\label{qChemistry}

Computational chemistry employs simulations to study properties of molecules or to predict unknown chemical phenomena. Indeed, analytical solutions for quantum many-body problems are available only for the simplest systems, thus making simulations necessary and heavily based on classical techniques such as the Born-Oppenheimer or Hartree-Fock approximations, the density-functional theory or even machine learning \cite{Olson17}. Notwithstanding, when modelling the quantum nature of highly correlated many-body systems, it became clear that the only solution to counteract the exponential increase in computational resources, the so-called \textit{curse of dimensionality}, was to exploit other quantum systems for information processing \cite{Feynman82}. The belief in this quantum approach was indeed supported in the 90's by the first efficient quantum algorithms \cite{Nielsen_Chuang}, which were rapidly followed by breakthrough discoveries applicable to quantum chemistry \cite{Lloyd96, Abrams99}. The algorithm developed in 2005 by Aspuru-Guzik \textit{et al.} \cite{AspuruGuzik05} offered an exponential speed-up in computational resources, scaling linearly in the number of qubits and polynomially in the number of gates. Today there exist numerous algorithms for simulating quantum chemistry, for which a complete review goes beyond the scope of this section. The reader interestered in a detailed and comprehensive excursus of the state of the art may find Ref. \cite{Olson17} a useful resource, while Refs. \cite{AspuruGuzik12, Noh16} may be suitable for a more experiments-oriented overview. In the following, we will briefly describe the latest achievements of quantum chemistry simulations on photonic platforms.

Photonic technologies provide an effective platform for quantum simulation, thanks to the single photons' low decoherence, speed and controllability that we discussed in the previous chapters. First experimental demonstrations in bulk optics were reported in 2009 for the simulation of anyons, fractional-statistics particles responsible for the fractional quantum Hall effect \cite{Lu09, Pachos09} and for the calculation of the energy spectrum of a hydrogen molecule \cite{Lanyon10}. The approach adopted in the latter work consisted in encoding the state of the molecule on single-photon polarization qubits, simulating its evolution in the Born-Oppenheimer approximation and estimating the energy $E$ of its eigenstates $\ket{\psi}$ using the quantum phase estimation (QPE) algorithm \cite{Abrams99}, since $ e^{i E t / \hbar} \ket{\psi} = e^{i 2 \pi \phi} \ket{\psi} $. In this case, with an appropriate choice of the basis, the Hamiltonian matrix reduces to two 2$\times$2 blocks (plus two 1$\times$1 blocks), allowing a map between subspaces and single qubit states. Since the operators act on single qubits, it was possible to perform QPE separately on the polarization qubit of the quantum register, accessing the outcome via the control entangled photon \cite{Lanyon10}. In 2010, Kaltenbaek \textit{et al.} generated a photonic valence-bond-solid state, the gapped ground state of a two-body Hamiltonian on a spin-1 chain, as a useful resource for the implementation of single-qubit quantum logic gates \cite{Kaltenbaek10}. In 2011, Ma \textit{et al.} used two polarization entangled photon pairs to study the process of frustration in a tetramer, a system with four spin-1/2 particles, observing the transition between localized and resonating-valence-bond states at varying Heisenberg interaction strength \cite{Ma11, Ma14}. The analog simulation required measuring the output of tunable beam splitter injected with single-photon states from each pair. All states have been characterized by retrieving the total energy and the pairwise quantum correlations, whose values are conditioned by the quantum monogamy. In 2012, Kitagawa \textit{et al.} employed a discrete-time bulk quantum walk to demonstrate the topological protection of bound states in both static and dynamic scenarios, a useful tool that finds application for instance in quantum computation. Topological transitions have been also the focus of a recent study carried out with orbital angular momentum encoding \cite{Cardano16}.

Recently the development of integrated photonic circuits for quantum walks (QW) has prompted the realization of various quantum simulations (see Section \ref{sec:QW}). Among discrete-time QWs, for instance, we have discussed the experimental investigation on the transition between fermionic and bosonic states \cite{Sansoni12}, enabled by polarization-encoded symmetric (bosonic) or anti-symmetric (fermionic) wave functions. Very recently, a new quantum approach based on variational methods and phase estimation was introduced \cite{Santagati18}, and experimentally tested, to approximate eigenvalues for ground and excited states. Further results were reported also with continuous-time QWs, with the observation of two-photon quantum correlations \cite{Peruzzo10, Lahini12, Matthews13, Poulios14}, quantum Rabi model \cite{Crespi12}, floquet topological insulators \cite{Rechtsman13}, growth of entanglement in spin chains \cite{Pitsios17}, Majorana dynamics \cite{Keil15}, Fano interference \cite{Crespi15fano} and Bloch oscillations \cite{Lebugle15}.

We conclude this section mentioning a very recent result in quantum simulation, which connects it to that of \textsc{BosonSampling} (see Section \ref{sec:BosonSampling}). On one hand, in fact, quantum simulations were conceived to surpass the capabilities of classical computers. On the other hand, Aaronson and Arkhipov have shown \cite{AA} that \textsc{BosonSampling}, i.e. the simulation of many-boson statistics, is indeed capable to provide a concrete evidence of this computational advantage. Within this framework, in 2015 Huh \textit{et al.} found the first application of \textsc{BosonSampling} for quantum simulation to evaluate molecular vibronic spectra \cite{Huh15, Peropadre16, GonzalezAlonso16bs}.

\section*{Discussion}

Photonic quantum technologies provide a promising platform for researches and applications in several contexts. This review attempted to gather their most recent advances, to provide the reader with a unified framework for the various ingredients. General aspects addressed in this manuscript are the generation, manipulation and detection of single-photon states from the technological perspective, as well as the fundamental theoretical tools developed for quantum communication and quantum simulation. The large effort devoted to these technologies is indeed testified also by the several achievements reported during the completion of this review, which make it challenging to stay up-to-date in this rapidly evolving field.

\section*{Acknowledgments}
This work was supported by the ERC-Starting Grant 3D-QUEST (3D-Quantum Integrated Optical Simulation; grant agreement no.307783; http://www.3dquest.eu)  and by the H2020-FETPROACT-2014 Grant QUCHIP (Quantum Simulation on a Photonic Chip; grant agreement no.641039; http://www.quchip.eu).

\section*{References}


\providecommand{\newblock}{}

\end{document}